\documentclass[12pt]{spieman}  
\usepackage{amsmath,amsfonts,amssymb}
\usepackage{graphicx}
\usepackage{setspace}
\usepackage{tocloft}
\usepackage{lineno}

\DeclareMathOperator*{\argmin}{arg\,min}
\newcommand{\edited}[1]{\textcolor{black}{#1}}

\title{The Holographic Dispersed Fringe Sensors (HDFS): phasing the Giant Magellan Telescope}

\author[a,*,$\dagger$]{Sebastiaan Y. Haffert}
\author[a]{Laird M. Close}
\author[a,b]{Alexander D. Hedglen}
\author[a]{Jared R. Males}
\author[a,b]{Maggie Kautz}
\author[c]{Antonin H. Bouchez}
\author[c]{Richard Demers}
\author[c]{Fernando Quir\'{o}s-Pacheco}
\author[c]{Breann N. Sitarski}
\author[a]{Kyle Van Gorkom}
\author[a]{Joseph D. Long}
\author[a,b,d,e]{Olivier Guyon}
\author[f]{Lauren Schatz}
\author[f]{Kelsey Miller}
\author[a,b]{Jennifer Lumbres}
\author[a,b]{Alex Rodack}
\author[a,b]{Justin M. Knight}

\affil[a]{University of Arizona, Steward Observatory, Tucson, Arizona, United States}
\affil[b]{Wyant College of Optical Science, University of Arizona, 1630 E University Blvd, Tucson, AZ 85719, USA}
\affil[c]{GMTO Corp., Pasadena, CA, USA}
\affil[d]{Astrobiology Center, National Institutes of Natural Sciences, 2-21-1 Osawa, Mitaka, Tokyo, JAPAN}
\affil[e]{National Astronomical Observatory of Japan, Subaru Telescope, National Institutes of Natural Sciences, Hilo, HI 96720, USA}
\affil[f]{Kirtland Air Force Base, Air Force Research Laboratory, Albuquerque, NM, USA}

\cftpagenumbersoff{figure}
\cftpagenumbersoff{table} 
\begin{document} 
\maketitle

\begin{abstract}
The next generation of Giant Segmented Mirror Telescopes (GSMT) will have large gaps between the segments either caused by the shadow of the mechanical structure of the secondary mirror (E-ELT and TMT) or intrinsically by design (GMT). These gaps are large enough to fragment the aperture into independent segments that are separated by more than the typical Fried parameter. This creates piston and petals modes that are not well sensed by conventional wavefront sensors such as the Shack-Hartmann wavefront sensor or the pyramid wavefront sensor. We propose to use a new optical device, the Holographic Dispersed Fringe Sensor (HDFS), to sense and control these petal/piston modes. The HDFS uses a single pupil-plane hologram to interfere the segments onto different spatial locations in the focal plane. Numerical simulations show that the HDFS is very efficient and that it reaches a differential piston rms smaller than 10 nm for GMT/E-ELT/TMT for guide stars up to 13th J+H band magnitude. The HDFS has also been validated in the lab with MagAO-X and HCAT, the GMT phasing testbed. The lab experiments reached 5 nm rms piston error on the Magellan telescope \edited{aperture}. The HDFS also reached 50 nm rms of piston error on a segmented GMT-like aperture while the pyramid wavefront sensor was compensating simulated atmosphere under median seeing conditions. The simulations and lab results demonstrate the HDFS as an excellent piston sensor for the GMT. We find that the combination of a pyramid slope sensor with a HDFS piston sensor is a powerful architecture for the GMT.
\end{abstract}

\keywords{adaptive optics, giant segmented mirror telescopes, phasing, wavefront sensing}

{\noindent \footnotesize\textbf{*}Address all correspondence to Sebastiaan Y. Haffert \linkable{shaffert@arizona.edu} \newline $\dagger$ NASA Hubble Fellow}

\begin{spacing}{2}   

\section{Introduction}
The size of the next generation of ground-based telescopes prohibits the use of monolithic mirrors. To still achieve larger diameter telescopes, the primary mirror is built up from an array of smaller segmented mirrors. There are currently two different approaches that are used to create giant apertures. The first is to use many small mirrors to assemble a single large diameter mirror. This method that has been chosen for the European Extremely Large Telescope (E-ELT) which will have 792 segments and for the Thirty Meter Telescope (TMT) with 492 segments. The Giant Magellan Telescope (GMT) chose a different approach, and uses 7 large (8.4 m) mirrors to construct the 25.4-meter aperture.

\begin{figure*}[htbp]
 \centering
 \includegraphics[width=\textwidth]{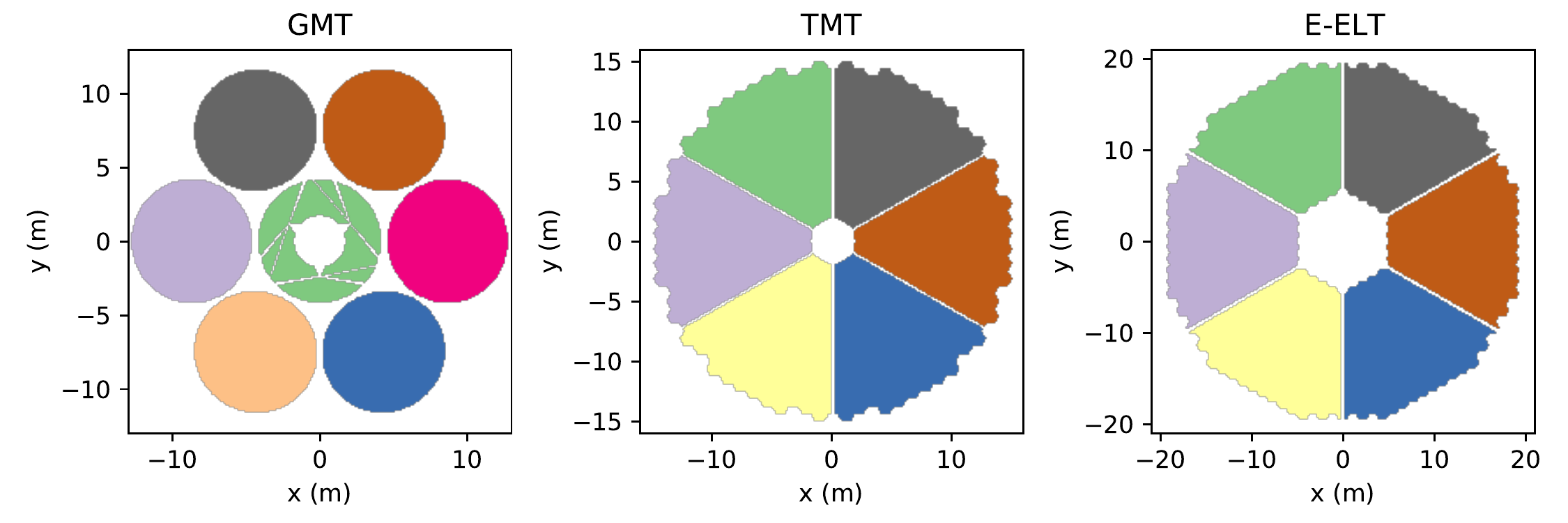}
 \caption{The aperture functions for the GMT, TMT, and E-ELT. The colors highlight the segment and petal modes that are difficult to sense with the PWFS. Each petal of the TMT and E-ELT consist of many smaller segmented mirrors that are phased together.}
 \label{fig:aperture_modes}
\end{figure*}

All the Giant Segmented Mirror Telescopes (GSMT) will use adaptive optics to reach diffraction-limited imaging. However, diffraction-limited quality is not always reached even with AO on such large segmented apertures. The mechanical support structure of the secondary mirror creates thick shadows in the pupil, which will fragment the pupil in several disjoint segments \edited{for the ELT and TMT. While} in the case of the GMT, the pupil is intrinsically segmented. Figure \ref{fig:aperture_modes} shows the segment modes for each of the telescopes. \edited{In this work we refer to the phase differences between these segments when we are talking about differential phase or piston errors. For the ELT these modes are often called petal modes. The phasing of the actual ($\sim 1$m sized) individual mirror segments of the ELT and TMT that make up the petal modes are not considered in this work.} Wavefront sensors that measure the local slope (e.g. the SHWFS), cannot measure differential piston between the pupil fragments because the distance between the segments is larger than the atmospheric coherence length. Many instruments have \edited{therefore} switched to the Pyramid Wavefront Sensor (PWFS), because it is more sensitive \cite{ragazzoni1999sensitivity} and the differential piston can be sensed in certain situations \cite{esposito2005pyramid}. The sensitivity of the PWFS to differential piston depends strongly on the modulation radius \cite{esposito2002segmented}. An unmodulated PWFS behaves similarly to an interferometer \cite{verinaud2004nature} and can therefore sense differential piston. And just like any interferometer, the PWFS has phase wrapping issues and can only sense the piston up to a multiple of $\lambda/2$ \cite{pinna2006phase}. In practice, the PWFS is never used on-sky unmodulated due to its limited dynamic range. When the PWFS is modulated, the response becomes similar to a gradient (or slope) sensor, which means that the modulated PWFS loses sensitivity to the differential piston \cite{esposito2002segmented, bertrou2020petalometry}. Both issues lead to strong differential piston residuals, which decreases the image quality significantly. More recent numerical simulations have shown that partial AO correction lead to increased cross-talk of higher-order modes into the segment piston modes \cite{bertrou2021confusion}. This makes it very difficult for the PWFS to control segment piston when operating under atmospheric turbulence conditions.

The phasing problem of "small gap ($<$1 cm)" segmented telescopes have been extensively studied, which led to the successful implementation of the phasing sensors on the two Keck telescopes \cite{chanan1998phasing,chanan2000phasing}. The phase sensor uses small sub-apertures that cover the edges between uncorrelated domains. Each sub-aperture creates an interference fringe from which the differential piston between the domains can be derived. When the differential piston changes, the fringe pattern moves. Just measuring the interference fringe is not enough, because there is still a phase wrapping ambiguity. The ambiguity can be resolved by dispersing the fringe. The dispersed fringe sensor (DFS) has been extensively tested in the lab \cite{shi2004experimental} and on the sky at the Magellan Telescope \cite{kopon2016sky}, and has been selected as the coarse phasing sensor for JWST\cite{shi2008nircam,acton2012wavefront} and has seen recent success controlling and correcting the differential piston at the Large Binocular Telescope for nulling interferometry \cite{spalding2018towards}.

The success of DFS has made it the current baseline sensor for differential piston for the GMT \cite{van2016dispersed, mcleod2018acquisition}. However, the small ($1.0\times1.5$m) sub-apertures of the DFS measure the differential piston very locally. This means that low-order modes can easily mimic differential piston signal. To circumvent this issue, at least 3 sensors per (8.4 m) mirror segment were necessary to compensate for segment tip-tilt. If higher-order modes are also influencing the reconstruction, even more sensors will be necessary. Secondly, the relative contribution of the thermal background increases for smaller sub-apertures because the diffraction-limited PSF size covers a larger patch of sky, which limits the guide star magnitude. And finally, each sub-aperture needs to be dispersed and imaged onto a detector. This makes for a complicated opto-mechanical instrument if many sub-apertures are required.

We propose to use a new optical device which we call the Holographic Dispersed Fringe Sensor (HDFS) as an alternative to the DFS. \edited{The HDFS will be used as a second channel wavefront sensor after a conventional AO system and it will purely measure the phasing of the segments in case of the GMT and the differetial phase of the petal modes for the ELT or TMT.} The HDFS uses holography to selectively interfere the entire individual pupil segments with each other. This holography is a much more elegant approach than trying to use mirrors and lenses to reach the same goal. The advances in manufacturing of phase holograms make it possible to create complex functions \cite{doelman2017patterned}. Multiplexed holograms are holograms that combine several functions into a single hologram, and, they are a prime example of a complex phase function. In past designs, multiplexed holograms have been used to create focal plane wavefront sensors \cite{wilby2017coronagraphic, wilby2016designing}, holographic aperture masks \cite{doelman2018multiplexed, doelman2021first}, and lateral shear interferometers \cite{por2021multiple}. We can integrate all the required DFS in a single hologram with the multiplexing approach. This results in a single pupil plane optic that can be used in any instrument with an accessible pupil plane. In Section 2, we describe the principles behind the HDFS and its design considerations and the piston reconstruction algorithm. Section 3 explores the performance of the HDFS with numerical simulations for the GMT and E-ELT. A HDFS hologram has also been manufactured and tested in the lab under realistic conditions. The lab results are shown in Section 4, and the conclusions are drawn in Section 5.

\section{The Holographic Dispersed Fringe Sensor}
\subsection{The Dispersed Fringe Sensor}
The HDFS makes use of diffraction and interference to encode the differential piston into a measurable signal. When two apertures interfere in focal plane, they create a fringe pattern. This fringe pattern modulates the individual Point Spread Function (PSF) of each aperture. The interference fringe can be described as
\begin{equation}
    I = |E_0|^2 + |E_1|^2 + 2|E_0||E_1|\cos\left(2\pi \alpha f_x + \frac{2\pi}{\lambda}\delta \right).
\end{equation}
Here, $E_0$ and $E_1$ are the electric fields of the \edited{first} and \edited{second} aperture, respectively. The apertures are separated by $\alpha$ and \edited{$f_x$ is the focal plane coordinate in spatial frequency space}. The differential piston, $\delta$, changes the modulation. The interference pattern reaches a peak when the total phase is equal to a multiple of $2\pi$. The phase shift of the fringe pattern is also wavelength dependent. This is a very important point, because if only monochromatic light is used, the fringe wraps at multiples of the wavelength (the classic problem of $2\pi$ phase wrapping in piston). This will lead to indistinguishable images and so any piston measurement can be incorrect by $\pm N \lambda$. However, if multiple wavelengths are imaged simultaneously, the fringe pattern will always be unique and the true differential piston can be recovered.

\begin{figure*}[htbp]
 \centering
 \includegraphics[width=\textwidth]{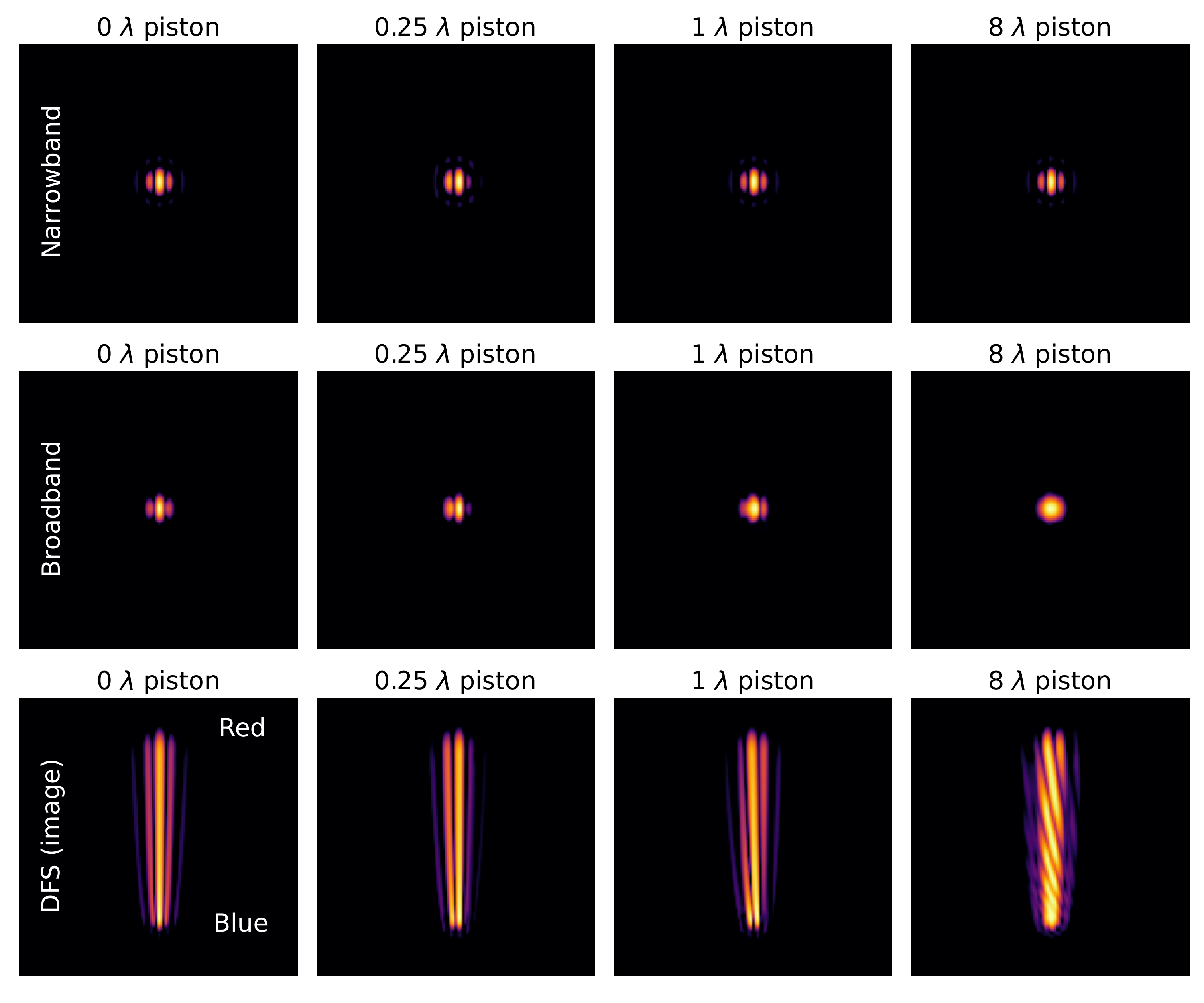}
 \caption{The interference pattern of two apertures in different bandpasses. From left to right the differential piston changes from $0\lambda$ to $8\lambda$. The top row show the fringe pattern of a narrowband source. For the middle row, the source has a broadband spectrum ($\Delta \lambda / \lambda=0.55$). And the bottom row shows the interference pattern when it is dispersed orthogonal to the baseline. The narrowband image is indistinguishable when the differential piston is a multiple of the wavelength. The broadband image can handle the wrapping, however when the piston is too large the apertures become incoherent. For the dispersed sensor, the differential piston causes a wrapping pattern that looks like a barber pole. The amount of wrapping is proportional to the amount of differential piston.}
 \label{fig:toy_model}
\end{figure*}

The (H)DFS works on basis of interference between two apertures, which means that the apertures need to add coherently. This is only possible when the differential piston between the apertures is smaller than the coherence length of the measurement. The coherence length of a light source is \edited{$L_{\mathrm{coh}} \approx \frac{\lambda^2}{\Delta \lambda}$}. When the differential piston becomes larger than the coherence length of a single spectral element, the fringes become incoherent and there will be no interference. This means that the dynamic range is approximately \edited{$L_{\mathrm{coh}}$}. Monochromatic sources, $\Delta \lambda \ll \lambda$, have the largest coherence length and therefore dynamic range, but as discussed before, monochromatic light has a wrapping issue. \edited{Another issue for monochromatic sources is that small spectral bandwidths also lead to low flux levels, which means that the measurement has low sensitivity.} Broadband sources, $\Delta \lambda \sim \lambda$, have no wrapping issues but have a very small dynamic range. This shows that dispersing the light to create multi-wavelength images is the only way to have the ability to uniquely determine the piston, while retaining a large enough dynamic range. The effects of the simple toy model that is discussed here are shown in Figure \ref{fig:toy_model}.

\subsection{Multiplexing the DFS with holograms}
For two apertures it is quite easy to create the DFS. \edited{An angularly dispersive element, such as a prism or grating,} has to be placed in the pupil plane and the fringe will be dispersed in the subsequent focal plane. The situation is not as simple for a segmented telescope aperture. For a segmented telescope the differential piston between all segments has to be measured and controlled. However, the differential piston can not be recovered from a single dispersed fringe due to symmetries in the telescope pupil. Symmetric apertures have repeated baselines that create fringes with the same frequency and orientation, leading to confusion of the differential piston. Therefore, the individual segments have to be interfered with each other in pairs at different spatial locations in the focal plane to measure the differential piston between pairs of segments. This was solved in the DFS for the GMT's active optics AGWS guider units by making many DFS from 12 prisms each with a different orientation that interfered 1.5m sections of the edges of each 8.4m segment \cite{kopon2016sky,kopon2017phasing}. All these DFS optics can be replaced by a single HDFS phase hologram.

The core concept of the HFDS is the diffraction grating. A grating diffracts an incoming beam of light into different diffraction orders ($m$), each having a position determined by the grating equation. The diffracted angle of a beam, $\theta_{out}$, that hits a grating, with a line period of $\Lambda$, at normal incidence is
\begin{equation}
    \sin{\theta_{out}} = m\frac{\lambda}{\Lambda}.
\end{equation}
If the outgoing angles are small, and there are $\beta$ periods across the pupil of diameter $D$, the grating equation becomes
\begin{equation}
    \theta_{out} = m \frac{\beta}{D}\lambda = m\alpha\frac{\lambda}{D}.
    \label{eq:grating}
\end{equation}
The grating creates a copy of the PSF at each diffraction order $m$. The shape of the grating groove determines the actual efficiency and phase of the PSF copy \cite{goodman2005introduction}. However, the position itself is only dependent on the grating period and the grating \edited{orientation (direction the grating lines run)}. This means that a copy of the PSF of a specific segment can be created by applying a grating on that segment. And, the interference fringe of a pair of segments can be created by applying identical gratings on both segments. All diffraction orders of the grating will then overlap and create fringes. An added benefit is that each fringe is dispersed because gratings are inherently dispersive (note the $\lambda$ dependence of Equation \ref{eq:grating}). The resolving power of the fringe is equal to the number of illuminated lines, $R=\alpha$. However, some segments will need to be interfered with multiple (typically 2) other segments to measure and constrain the differential piston across all segments. 

\begin{figure*}[htbp]
 \centering
 \includegraphics[width=\textwidth]{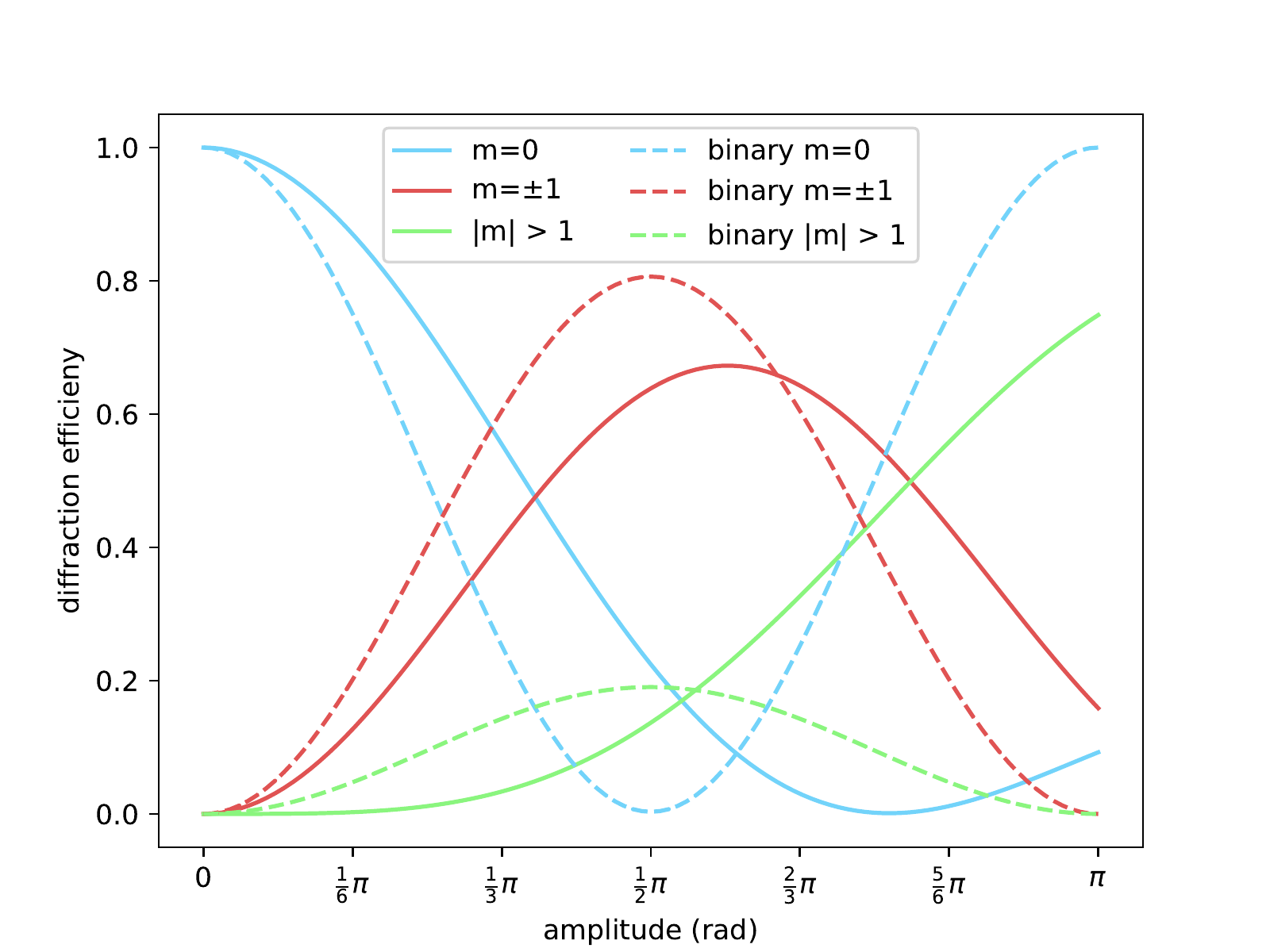}
 \caption{Diffraction efficiency as function of the grating amplitude for various diffraction orders. All the orders larger than $\pm$1 have been combined. The solid line shows the efficiency for the continuous grating and the dashed line shows the efficiency for the binary grating.}
 \label{fig:diffraction_efficiency}
\end{figure*}

Multiple gratings can be multiplexed on the same segment by adding the phase profiles of each grating\edited{, at the cost of diffraction efficiency}. The HDFS phase profile for segment $j$ across the pupil coordinates $(x,y)$ is then
\begin{equation}
    \phi_j = \sum_i A_{j} \cos{\left[2\pi \alpha_{ij}\left(x\cos{\psi_{ij}} - y\sin{\psi_{ij}}\right) \right]}.
\end{equation}
With, $\phi_j$ the phase profile for segment $j$, $A_{j}$ the aperture function of the segment, $\alpha_{ij}$ the grating period for segment $j$ and grating $i$, and $\psi_{ij}$ the orientation of the grating. However, this direct approach of multiplexing leads to less efficient gratings because there is more cross-talk between the different diffraction gratings, and because a lot of light is retained in the zeroth order. Blazed gratings have the highest first-order diffraction efficiency, most of the light is concentrated in the dispersed fringes instead of the zeroth order. The gratings in this work are blazed by making them binary. The binary hologram can be found by taking the sign of the phase profile from the direct multiplexing method,
\begin{equation}
    \bar{\phi}_{i} = \frac{\pi}{2} \mathrm{sgn}\left(\phi_i\right).
\end{equation}
Only the strength, i.e. the amplitude scaling, of the binary hologram has to be chosen. The grating amplitude should be chosen as $\pi/2$, if all light has to be diffracted in the dispersed fringes. The total phase shift between the two levels is then $\pi$, which is the maximum phase shift that can be applied due to the phase wrapping property of the electric field. The diffraction efficiency of both the continuous and binary holograms are compared for different amplitudes in Figure \ref{fig:diffraction_efficiency}. This figure shows that the peak efficiency of the binary pattern indeed occurs at $\pi/2$. The binary hologram is preferred over the continuous version because the binary hologram is not only more efficient (82\% versus 65\%), the other orders also have a much lower efficiency. This will lead to lower cross talk between the multiplexed holograms.

\subsection{Constraints on the HDFS parameters}
\begin{figure*}[htbp]
 \centering
 \includegraphics[width=\textwidth]{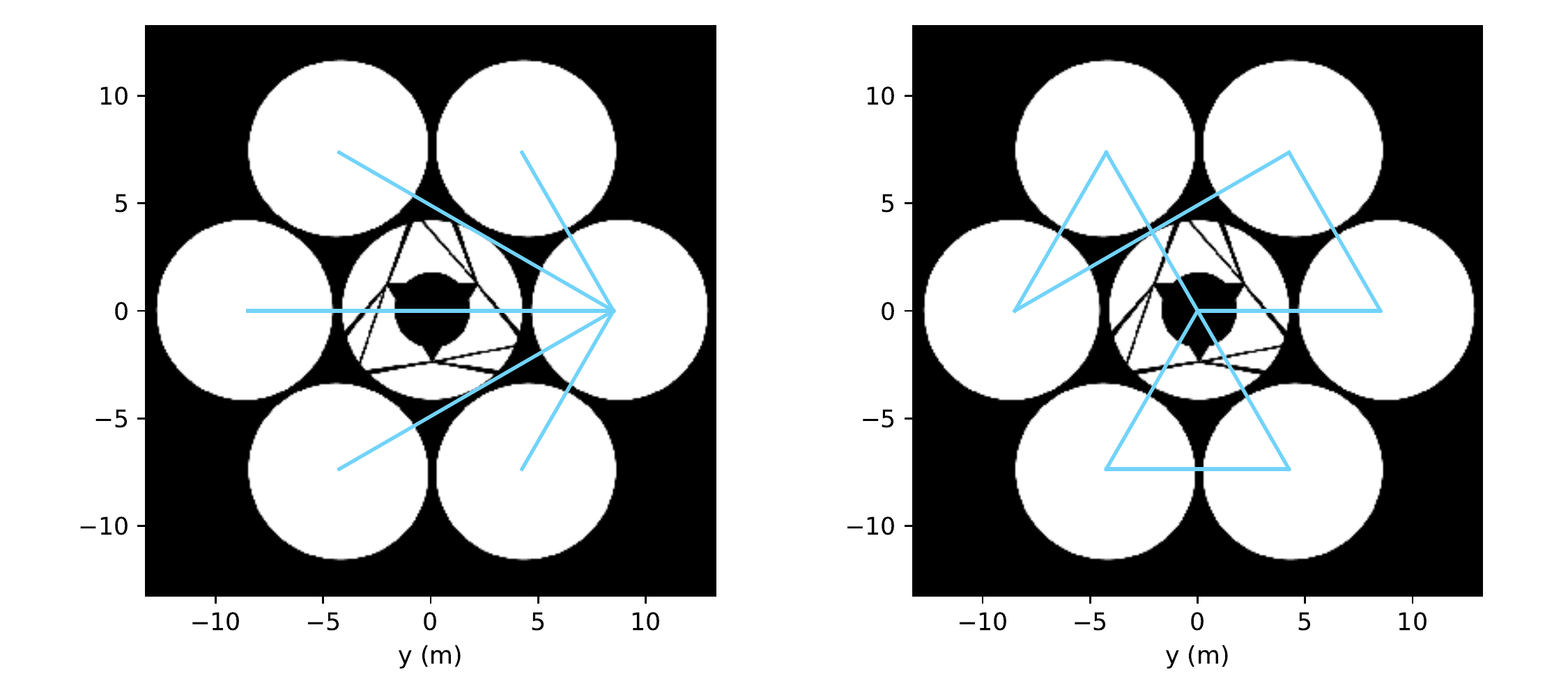}
 \caption{The GMT aperture overlaid with the connected baselines. The left figure shows the configuration where the right segment is the phase reference of all segments. The blue lines show the baselines for all the required fringes. The figure on the right shows the chain approach, where each segment is chained to two other segments. Each segment will always be connected to two other segments.}
 \label{fig:chain_design}
\end{figure*}

The visibility of the HDFS fringes should be maximized for the highest signal-to-noise. This can be achieved by giving the PSF copies of each segment equal power. A consequence of this choice is that the HDFS should not use a single segment as a reference for all other segments. If one segment is chosen as the reference for all other fringes, it will have many multiplexed gratings while the other segments only have a single grating. If there are $N$ segments, then that one reference segment will need to split its light over $N-1$ copies. For maximum visibility the power in the copies should be equal, which means that only $1/(N-1)$ of the power can be used for each measurement. This significantly lowers the signal-to-noise. Therefore, it is better to chain pairs of segments. The difference between the two approaches is shown schematically for the GMT aperture in Figure \ref{fig:chain_design}. Each segment has two multiplexed gratings in the chain approach. This leads to optimal photon efficiency because the amount of multiplexed gratings is equal for all segment, and therefore maximum visibility can be achieved with all power. The chain method is therefore highly preferable compared to the single reference segment method. \edited{However, this does not mean that the single reference segment method does not work. Its behavior is identical to the chained method, only its SNR at equal photon counts is lower.}

\begin{figure*}[htbp]
 \centering
 \includegraphics[width=\textwidth]{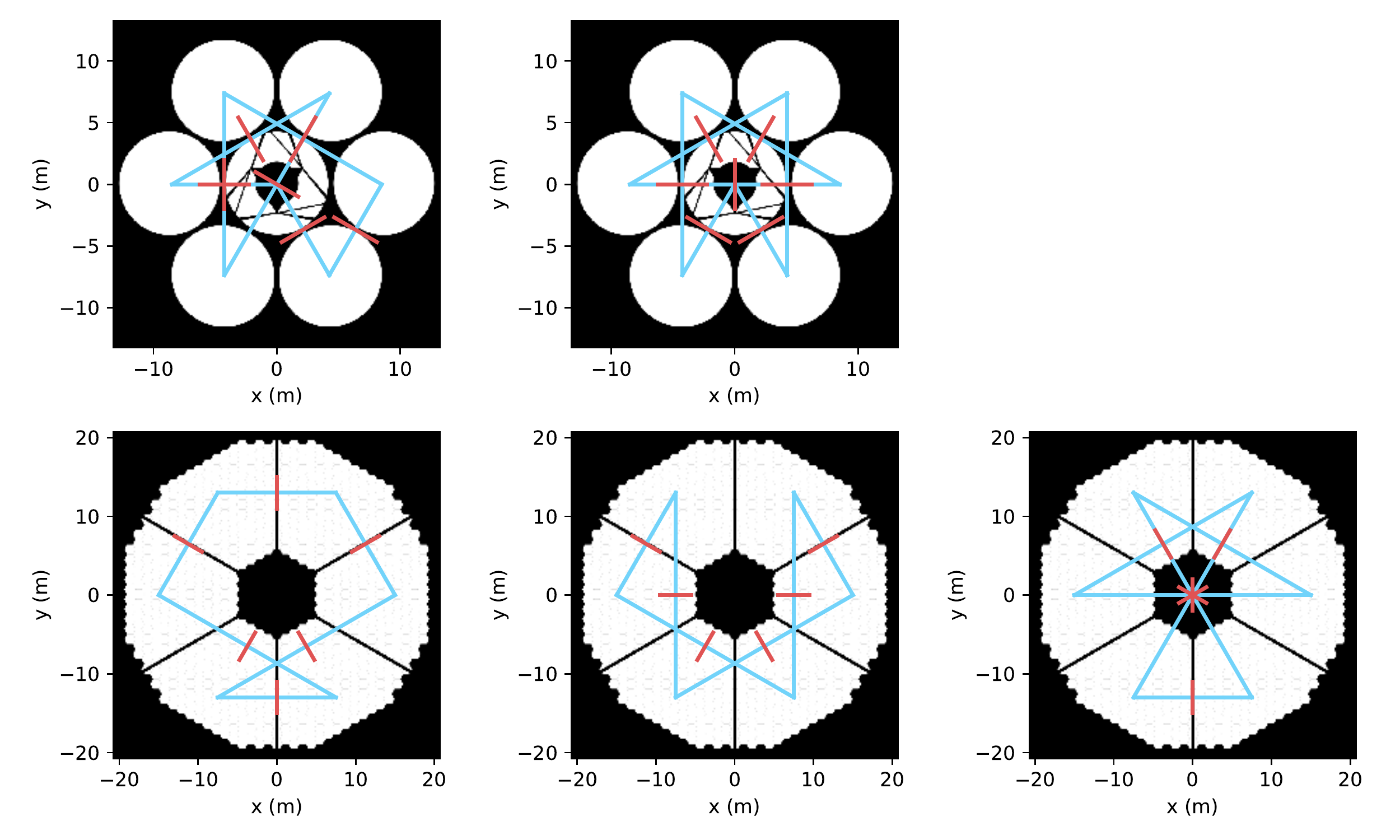}
 \caption{The top row shows the GMT aperture overlaid with the connected baselines. The E-ELT aperture overlaid with the connected baselines is shown on the bottom row. The blue lines show the baselines for each fringe, while the red lines show the direction of the dispersion. Due to redundancy of these pupils there is always 2 red lines that are parallel (the rest are unique). These two dispersion angles must be slightly tilted w.r.t. the perpendicular so as to not overlap. All 5 others are unique.}
 \label{fig:unique_chains}
\end{figure*}

Another constraint of the HDFS is that the fringes have to be dispersed near orthogonal to the baseline that connects the chosen segments. The interference fringe itself is parallel to the baseline. If the dispersion is also in that direction, the fringes will be smeared out. This constraint strongly limits the sequence of pairs that can be used for the HDSF in the chain method. If two segment pairs are chosen with the same baseline, the dispersed fringes of those two pairs will overlap. Therefore, the baselines have to be as unique as possible. Each chain method design can be described by the sequence in which the segments are chained together. It is computationally cheap to find a solution that contains only unique baselines by going through all permutations of the chain sequence. The direction of the dispersed fringe is then calculated for each baseline in the sequence. A solution is determined to be unique if all directions are different. However, no unique solution could be found for the GMT aperture with 7 segments, or the 6 segment E-ELT/TMT aperture. The next step is to allow a single repeated baseline. There are two solutions with one repeated baseline for the GMT aperture, and three solution for one repeated baseline for the E-ELT/TMT aperture. The solutions are show in Figure \ref{fig:unique_chains}. The overlapping baselines can be separated by offsetting the dispersion direction of the fringes with a small angle. The offset will lower the visibility slightly, which is not a major problem if the offset is not too large.

\begin{figure*}[htbp]
 \centering
 \includegraphics[width=\textwidth]{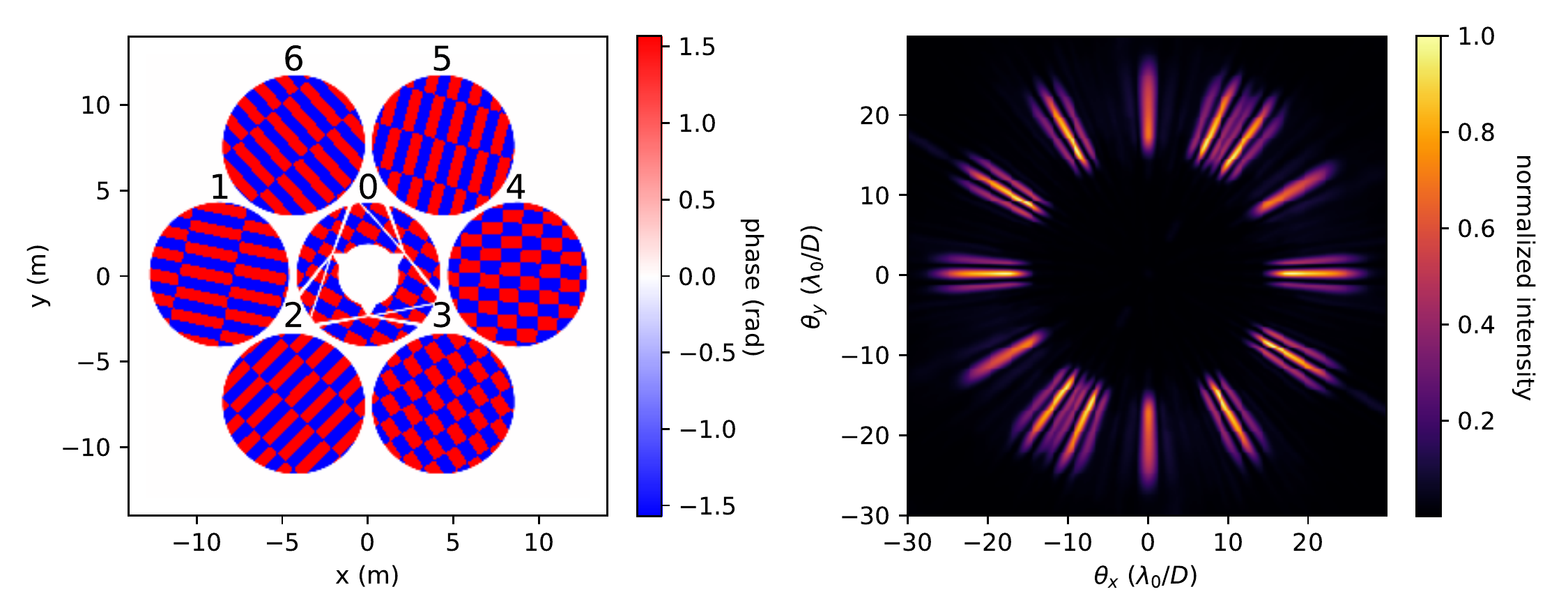}
 \caption{The left figure shows the HDFS pattern corresponding to GMT design 2 of Figure \ref{fig:unique_chains}. Each segment has its index number next to it. The image on the right shows the corresponding focal plane. Each grating creates two DFS "fringe" patterns on either side, one for the $m=1$ diffraction order and one for the $m=-1$. The binary hologram has a high efficiency because most of the light is concentrated into the $m=\pm1$ diffraction order. Hence, there is almost no light in the center.}
 \label{fig:hdfs_example}
\end{figure*}

An example of the HDFS pattern and its corresponding focal plane are shown in Figure \ref{fig:hdfs_example}. The design that is shown is the second GMT design from Figure \ref{fig:unique_chains}. An offset of $\pm \pi/30$ has been added to the repeated baselines to separate them in the focal plane. The carrier frequency has been set to $\alpha=20$, which means that theoretically this HDFS should be able to measure up to $20\lambda$ of differential piston.

\edited{The HDFS that is described in this section uses the full segments. However, it is also possible to select sub-apertures on the segments that will be interfered. This approach is similar to Sparse Aperture Masking, and, with the holographic technique we can choose to selectively interfere sub-apertures. This makes it easier to solve the unique baseline problem. Smaller apertures also have increased spatial coherence, which makes it easier use under strong turbulence conditions. The disadvantage is that the guide star magnitude limit will decrease as $D^4$ because of the larger PSF ($D^2$) and smaller aperture ($D^2$).}

\subsection{Reconstructing differential piston}
There are two approaches to measure the piston signal from the (H)DFS fringes. The Fourier method \cite{van2016dispersed}, or the template matching method \cite{chanan2000phasing}. In this work, we work with the template matching method. To reconstruct the piston signal it is possible to do a direct fit of the dispersed fringe,
\begin{equation}
    \hat{\delta} = \argmin_{\delta} (I - f(\delta))^2.
    \label{eq:sqdiff}
\end{equation}
The observed image $I$ is then compared with a model $f(\delta)$ generated for a particular amount of differential piston $\delta$. The best fitting piston signal $\hat{\delta}$ is found by minimizing the squared difference between the model and observation. However, this approach is computationally expensive because many wavelength channels and piston values have to be simulated. An alternative is to precompute the model images for a range of expected piston values. Expanding Equation \ref{eq:sqdiff} results in,
\begin{equation}
    \hat{\delta} = \argmin_{\delta} I^2 + f(\delta)^2 -2If(\delta)^2.
\end{equation}
The model and observation terms, $I^2$ and $f(\delta)^2$, are fixed because the observation has been done and the models have been computed. Therefore, the only term that varies between the various piston signals is the cross term $If(\delta)$. This term is equivalent to the cross-correlation of the observation and the precomputed model images. Minimization of the squared differences is therefore the same as maximizing the cross-correlation between the observation and models. The precision of this method is set by the size of the piston step that was used during the model computation. A 2nd-order polynomial is fit to the peak of the cross-correlation to get sub-sample precision. For the HDFS, we created a sequence of model templates for each fringe. If a new observation is done each fringe is extracted with a mask, and then compared to its own set of model fringes. This results in a vector of differential piston measurements between the chained segments.

The fringes only encode the differential piston between the segments that are chained together. An unwrapping method has to be applied to reconstruct the differential piston using a single segment as the phase reference for all other segments. A single segment can always be chosen as the absolute phase reference because the global piston does not matter, only the differential piston. The HDFS encodes the difference between segments in the fringes with a mixing matrix M. This effectively means there is a linear matrix that transforms the actual piston signals in a measurement vector, $\vec{S}=M\vec{\delta}$. The segments pairs for GMT Design 2 are \edited{$\{[0, 2], [2, 6], [6, 4], [4, 1], [1, 5], [5, 3], [3, 0]\}$}. The index numbers here correspond to the indices in Figure \ref{fig:hdfs_example}. If we assume that segment 0 is the absolute phase reference, we can express all segment piston relative to the piston of segment 0. This means that $\delta_{0}$ is always 0. In that case the mixing matrix is,
\begin{equation}
\vec{S} = \begin{bmatrix}
 -\delta_{2}\\ 
\delta_{2} - \delta_{6}\\ 
\delta_{6} - \delta_{4}\\ 
\delta_{4} - \delta_{1}\\ 
\delta_{1} - \delta_{5}\\ 
\delta_{5} - \delta_{3}\\ 
\delta_{3}\\ 
\end{bmatrix} = \begin{pmatrix}
 0 & -1 & 0 & 0 & 0 & 0\\ 
 0 & 1 & 0 & 0 & 0 & -1\\ 
 0 & 0 & 0 & -1 & 0 & 1\\ 
 -1 & 0 & 0 & 1 & 0 & 0 \\ 
 1 & 0 & 0 & 0 & -1 & 0 \\ 
 0 & 0 & -1 & 0 & 1 & 0 \\ 
 0 & 0 & 1 & 0 & 0 & 0 
\end{pmatrix} \begin{bmatrix}
\delta_{1}\\ 
\delta_{2}\\ 
\delta_{3}\\ 
\delta_{4}\\ 
\delta_{5}\\ 
\delta_{6}\\ 
\end{bmatrix}= M\vec{\delta}
\end{equation}
The pseudo-inverse of the mixing matrix is unwrapping matrix that allow us to reconstruct the differential piston signals. So, $\vec{\delta} = M^{\dagger} \vec{S}$. The  ${\dagger}$ operator is used to indicate a pseudo-inverse.


\subsection{Manufacturing of the HDFS}
The HDFS is a phase pattern that needs to work over a large bandwidth ($\Delta \lambda / \lambda=$ 50\%). Manufacturing achromatic phase patterns is difficult with conventional etching techniques. However, they can be created by using the geometric phase properties of polarization optics. When circular polarized light goes through a half-wave retarder, it will accrue a phase that depends on the angle of the fast-axis and the state of polarization \cite{escuti2016controlling}. This is called the geometric phase or the Pancharatnam-Berry phase \cite{pancharatnam1956generalized, berry1987adiabatic}. The amount of phase delay is, \begin{equation}
\phi = \pm 2 \theta.
\end{equation}
With $\phi$ as the amount of phase for a fast-axis angle of $\theta$. Geometric phase applies opposite phase to left and right circular polarized light. This has a consequence for the HDFS pattern. The left and right circular polarized light fringes will have opposite phase, which makes them incoherent to any potential cross-talk from the zeroth-order PSF \cite{bos2020vector}. This is an important property because the zeroth-order PSF can scatter light at the position of the dispersed fringes due to high-frequency wavefront errors. Such errors can come from atmospheric turbulence or even from polishing errors on the optics. However, the geometric phase HDFS cleanly decouples such interference terms, which makes geometric phase the prime choice to implement the HDFS.

\section{Simulations of the HDFS}
In this section, the properties of the HDFS for the GMT and E-ELT are explored through numerical simulations. All simulations are performed with the High Contrast Imaging for Python (HCIPy) package \cite{por2018high}. \edited{In these simulations the HDFS is used as a second channel wavefront sensor after a conventional AO system. This means that we assume that the AO system is delivering AO corrected segments (GMT) or petals (ELT/TMT). We note that slope sensors like the SHWFS or PWFS can flatten the wavefront over individual segments, even if there is significant piston error between these segments \cite{hedglen2022hcat}. So an architecture of a PWFS (slope sensor) followed by a HDFS (phase/petal sensor) can be a powerful solution to GMT/ELT wavefront sensing.}

\subsection{Measurement noise behavior}
The HDFS interferes the full pupils with each other, instead of small segments like the normal DFS. This provides a significant gain in sensitivity. For bright targets, there is more flux because of the larger utilized aperture. For example, the HDFS's 8.4m apertures collect 18x more light than the GMTs slow 1.5x1.0m, DFS apertures in its off-axis, seeing-limited, active optic AGWS guiders. For fainter targets there is another benefit. The limiting noise source for (N)IR wavelengths is the thermal background, which can come from either the sky or the telescope and instrument. A larger aperture will concentrate the PSF into a smaller area because the diffraction-limit decreases with increasing aperture. For example, an advantage of an additional 36x smaller area of the HDFS over the AGWS DFS. This reduces the contribution of background noise to the reconstruction error. The parameters of the simulation are shown in Table \ref{tab:photon_noise}. 

\begin{table}[]
\caption{The parameters of the HDFS simulations.}
\centering
\begin{tabular}{lll}
\hline
\textbf{HDFS Parameter} & \textbf{Value} & \textbf{Comment} \\ \hline \hline
Spectral bandwidth & 1125 nm to 1825 nm & covers J+H band \\ \hline
$\alpha$ & 75 lines &  \\ \hline
Integration time & 0.001 s & 1 ms \\ \hline
Throughput & 0.35 & QE of 0.8 and 0.45 instrument throughput \\ \hline
Magnitude zeropoint & $1.5\cdot10^{12}$ photons s$^{-1}$& averaged over J+H \\ \hline
Background & 14.6 mag arcsec${^{-2}}$ & averaged over J+H \\ \hline
read-noise & 0.4 e- & IR APD detector \\\hline
\end{tabular}
\label{tab:photon_noise}
\end{table}

\begin{figure*}[htbp]
 \centering
 \includegraphics[width=\textwidth]{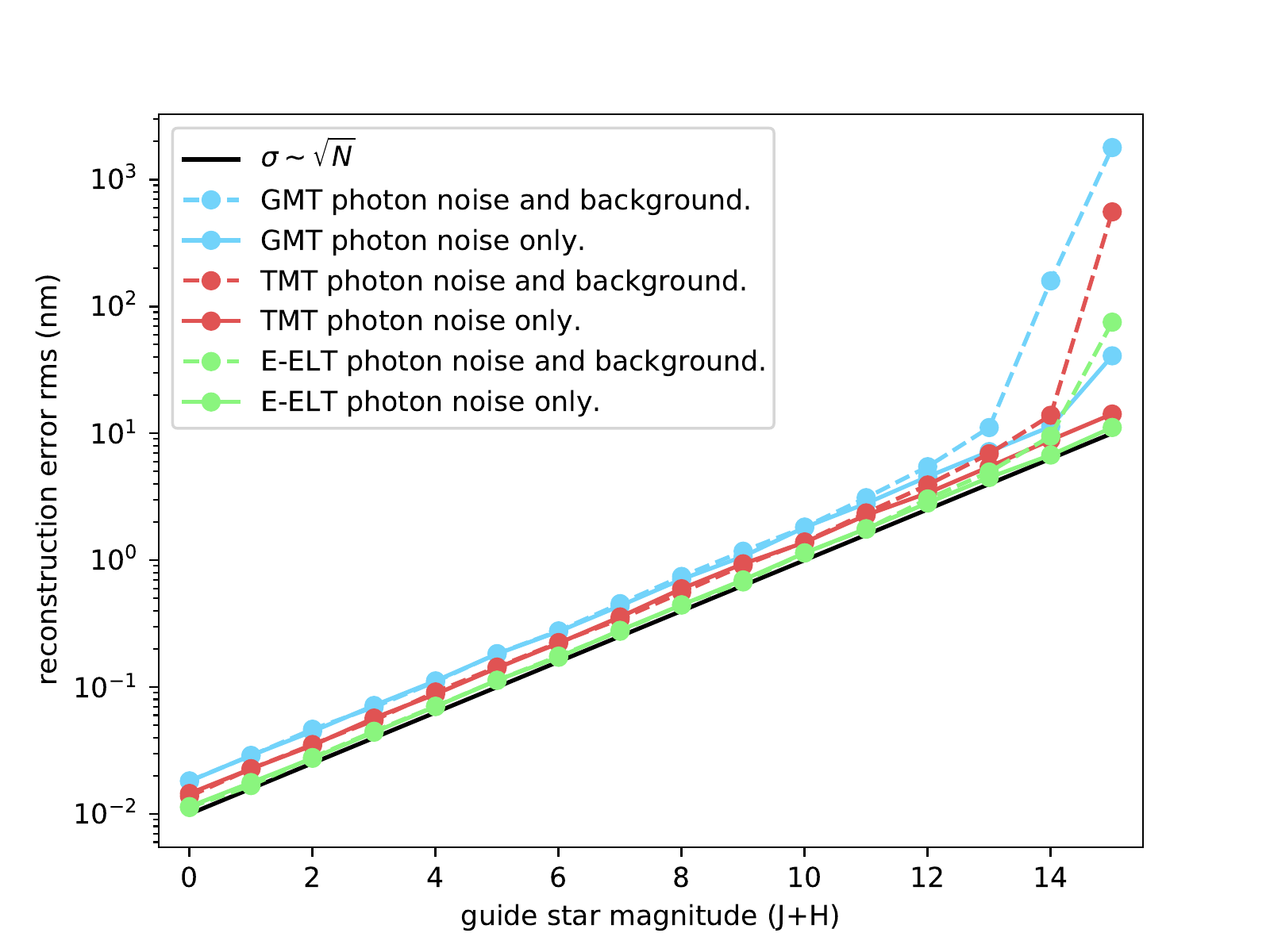}
 \caption{Reconstruction error as a function of guide star magnitude for an HDFS running at 1 kHz. The curves are shown for all three GSMT. The solid lines only include photon-noise errors, while the dashed lines also include background noise. The E-ELT performs best because it has the largest aperture, while the GMT has the worst performance because it has the smallest aperture. The difference is strongest at the fainter end, where the observations become background-noise limited.}
 \label{fig:photon_noise}
\end{figure*}

Both effects are visible in Figure \ref{fig:photon_noise}. The larger aperture of the E-ELT shows smaller reconstruction errors as compared to the other telescopes. And in the faint source limit, the reconstruction quality of the GMT degrades faster than the error of the E-ELT. The gain in sensitivity compared to the DFS is more apparent. The DFS reached an a reconstruction quality of 85 nm rms at 13.5 magnitude in simulations \cite{van2016dispersed}, where the HDFS reaches a reconstruction quality of 50 nm rms at the same guide star magnitude. However, the DFS simulation was with only the J-band and with an integration time of 10 ms. The simulations presented here for the HDFS use both the J and H-band, however an integration time of 1 ms is used. This shows that the HDFS has a higher sensitivity after correcting for the bandwidth and the integration. The HDFS can reconstruct piston errors down to an rms of 50 nm or smaller for the E-ELT, even all the way down to 15th J+H magnitude. Also, a cold field stop or a spatial filter in the focal plane of the HDFS channel could significantly reduce the background further.

\subsection{Low order sensitivity}
\begin{figure*}[htbp]
 \centering
 \includegraphics[width=\textwidth]{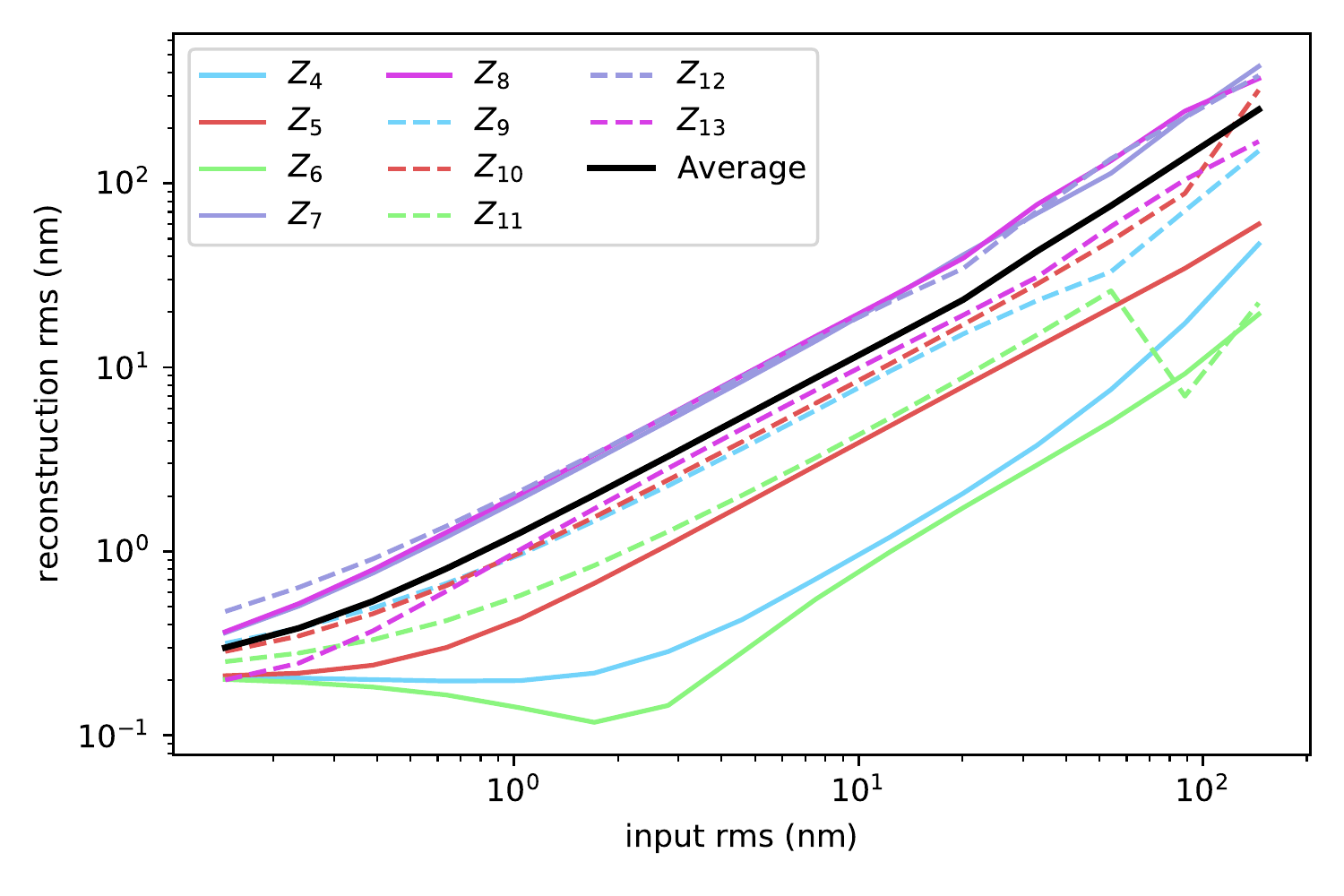}
 \caption{Reconstruction quality as function of input rms wavefront error for various Zernike modes. Tip and tilt are not shown because the Fourier-based approach is tip/tilt insensitive. The black line shows the reconstruction error for a uniformly distributed wavefront error containing all the low-order modes. The piston reconstruction error depends linearly (slope of 1 on a log-log plot) on the input rms wavefront error.} \label{fig:low_order_stability}
\end{figure*}

The HDFS is a focal plane wavefront sensor, which means that it is susceptible to cross-talk from low-order aberrations. Medium- to high-order aberrations do not influence the HDFS because only wavefront errors that influence the fringe shape matter. Medium to high-order wavefront errors create speckles far away from the fringes. This reduces their influence significantly as will be shown in the next section. Therefore, only low-order aberrations matter for the HDFS. Only the GMT HDFS design is analyzed for the low-order stability because the effects will be quite similar for the other apertures. The reconstruction rms as function of aberration rms is shown in Figure \ref{fig:low_order_stability}. Tip and tilt are excluded because these can be removed by post-processing the measurements. The reconstruction rms depends linearly on presence of low-order aberrations. The GMT design has the poorest performance in the presence of coma ($Z_7$ and $Z_8$) and secondary astimagtism ($Z_{12}$). The low-order aberrations need to be smaller than $\sim$40 nm rms for a reconstruction rms of 50 nm or smaller. These numbers have to be seen in the correct context. It is not the total wavefront error that has to be smaller than 40 nm rms, only the contribution of the low-order modes.

\subsection{Closed-loop behavior of the HDFS}
\begin{figure*}[htbp]
 \centering
 \includegraphics[width=\textwidth]{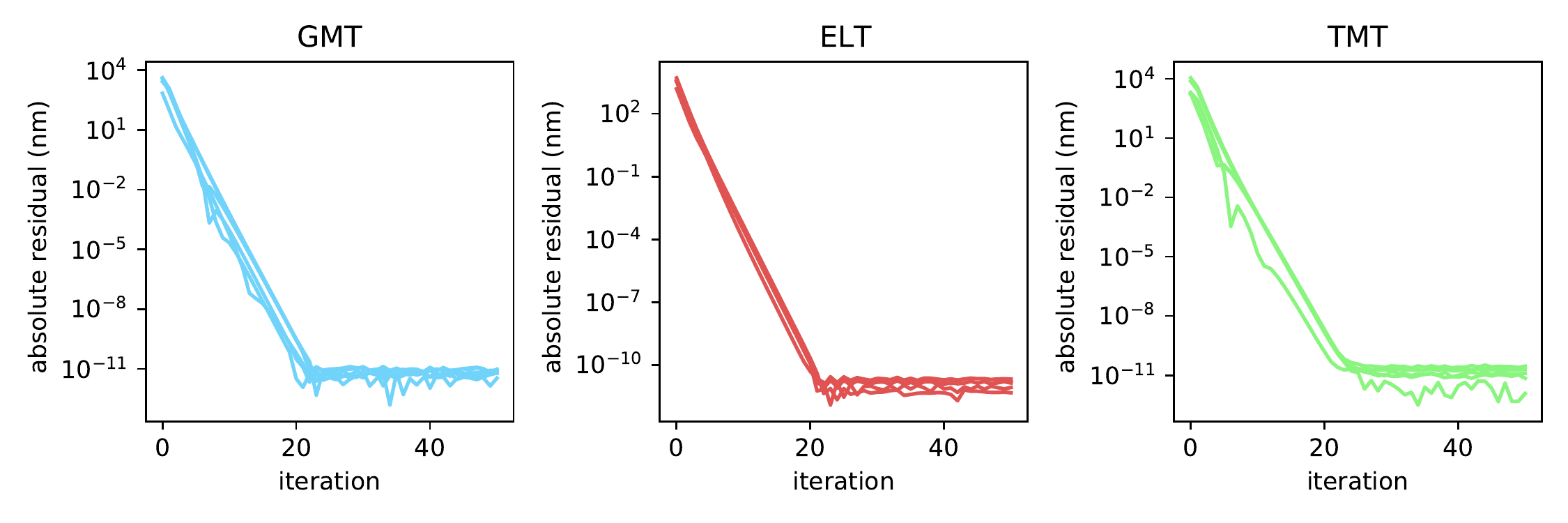}
 
 \caption{The piston residuals as function of time for the three GSMT. The HDFS reaches numerical zero after 20 iterations for each telescope.}
 \label{fig:closed_loop_hdfs}
\end{figure*}

The HDFS has good performance while running in closed-loop. It can reach numerical zero if no other sources of noise (photon noise or other wavefront errors) are present. The closed-loop residuals are shown in Figure \ref{fig:closed_loop_hdfs} for all three of the GSMT. This shows that the HDFS is in principle unbiased and can reach zero piston for all segments. The HDFS reaches $10^{-11}$ nm rms after 20 iterations and starting at several microns rms.

\begin{figure*}[htbp]
 \centering
 \includegraphics[width=\textwidth]{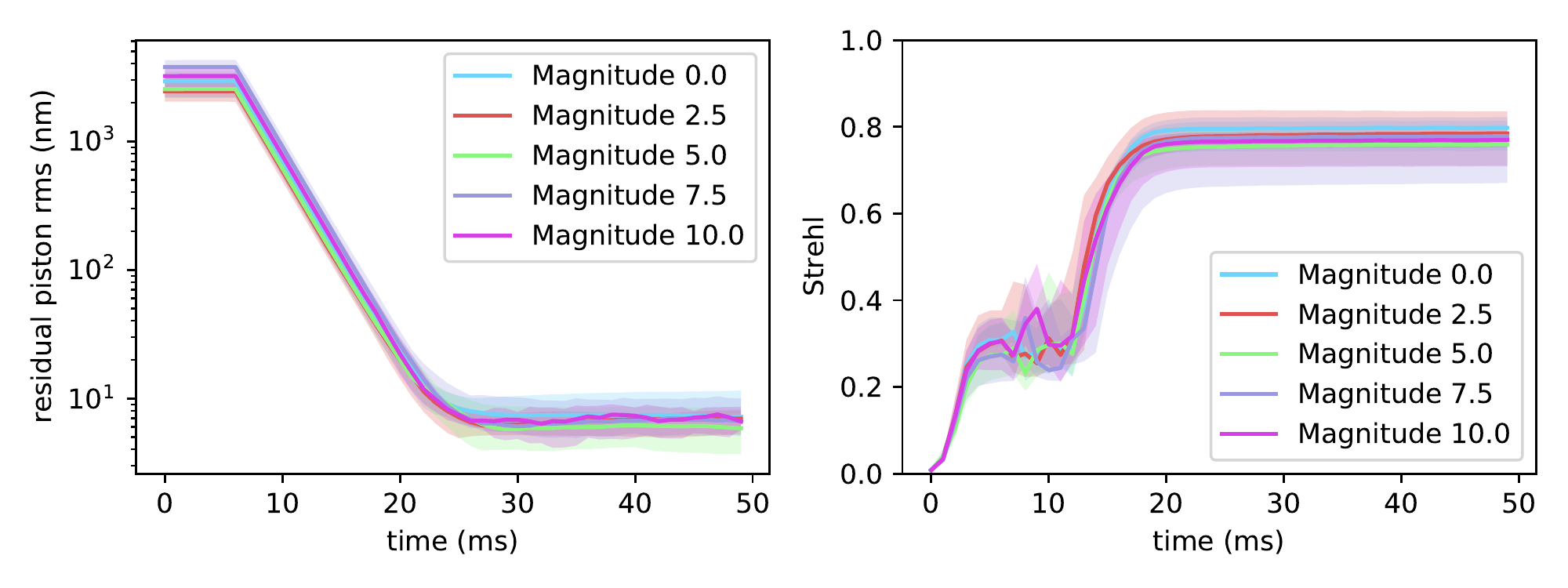}
 
 \caption{The left figure shows the piston rms as function of iteration for various stellar magnitudes. The figure on the right shows the corresponding Strehl ratios. The HDFS is turned on after iteration 5 (t=5 ms). The Strehl has large variation between iteration 5 to 10, which is due to unwrapping of the piston signals. This is also clear from the piston residuals themselves, because they are always decreasing. At iteration 10, the system is close to the white fringe and the Strehl increases monotonically.}
 \label{fig:closed_loop_hdfs_with_ao}
\end{figure*}

More importantly is the closed-loop behavior in the presence of atmospheric residual speckles. The effect of high-order speckles has been investigated for the GMT by using a simple AO system. The other telescopes are not investigated because the closed-loop behavior will be quite similar. We used a perfect linear wavefront sensor with a system lag of 1 frame, a gain of 0.5 and median seeing conditions (0.65" at Las Campanas observatory \cite{floyd2010seeing}). The wavefront sensing is performed by a linear projection on a grid of 64x64 Fourier modes (4K DM). The piston modes were measured and removed before the projection, otherwise the \edited{linear} wavefront sensor would try to control them. The piston modes are then injected again after the AO loop so that the HDFS would see the piston modes. Both the AO system and the HDFS are running at 1 kHz. The performance is tested for a range of magnitudes (0th to 10th). However, \edited{no photon noise is added to the wavefront sensor. The simulation shown here are done to test the robustness of the HDFS against high-order wavefront error residuals. Therefore, the effects of photon noise in the wavefront sensor are not relevant for these simulations.} So this is mainly a test for the HDFS's performance. The results of the closed-loop simulations are shown in Figure \ref{fig:closed_loop_hdfs_with_ao}. We ran ten realization for each magnitude to average out the statistical noise. The results show that the HDFS is never limited by the photon noise, because each stellar magnitude reaches roughly 5 nm rms. This is in line with the photon-noise sensitivity from Figure \ref{fig:photon_noise}. The rms is most likely limited by the closed-loop dynamics of the controller. The effects of lower Strehl from the AO system with increasing magnitude has not been explored and will be part of future work \edited{because this would require a detailed instrument concept and model}. These simulations show that the HDFS can be used within a AO system and that it is, in theory, possible to reach a piston rms smaller than 10 nm. 
\subsection{Amplitude errors}
\edited{Atmospheric scintillation did not have any significant effect on the HDFS. Scintillation causes amplitude variation on small scales due to the large aperture of the GSMT's. This means that scintillation effects the high-spatial frequencies the most. The HDFS uses the interference of the PSF cores and is mostly insensitive to high-spatial frequency errors. It is unlikely that scintillation will play a significant role.}

\edited{A stronger effect is expected from tranmission effects of the segments. The main impact of differential transmission is unequal power of the two PSFs from the segments that are interfered. This will lower the fringe visibility, but not the extracted signal. Figure \ref{fig:toy_model} shows that the signal is mainly encoded in the tilting and wrapping of the dispersed fringe. Changing the amplitude of the mirrors will not impact the orientation of the fringes only their depth. Therefore, only the SNR will be affected when the transmission of the segments change.}

\section{Lab results with P-HCAT and MagAO-X}
\begin{figure*}[htbp]
 \centering
 \includegraphics[width=\textwidth]{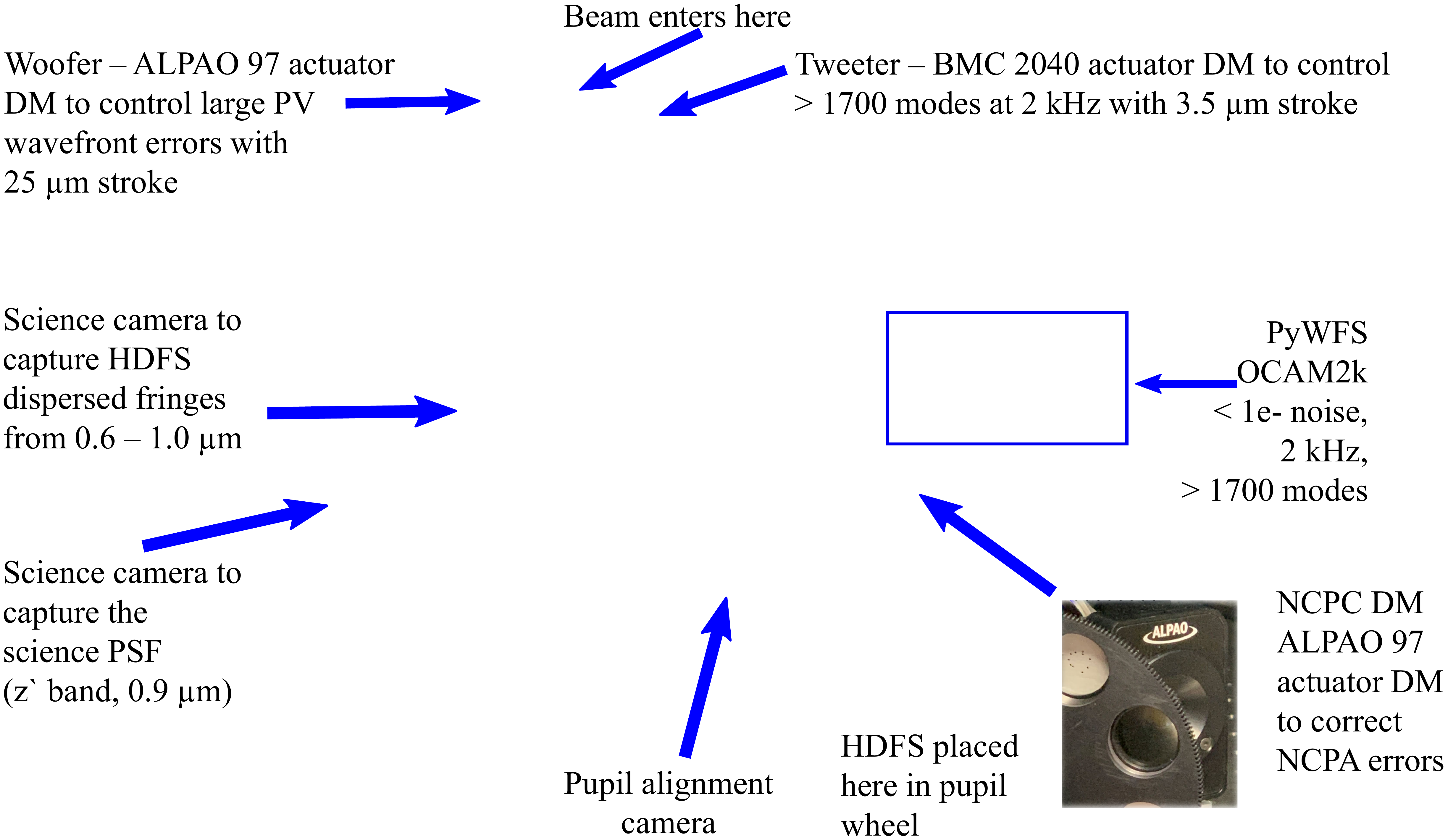}
 \caption{A CAD model of MagAO-X. The beam enters the top bench, and reflects of the Woofer and then the Tweeter DM. A periscope brings the beam down to the bottom bench, which accomodates the PWFS and science cameras. There is a third DM, the NCPC DM, which corrects low-order non-common path aberrations. Just after this DM is a pupil filter wheel (see the inset) that contains the HDFS phase hologram. The beam is then imaged onto the science detectors with a f/69 beam.}
 \label{fig:magaox}
\end{figure*}

\begin{figure*}[htbp]
 \centering
 \includegraphics[width=\textwidth]{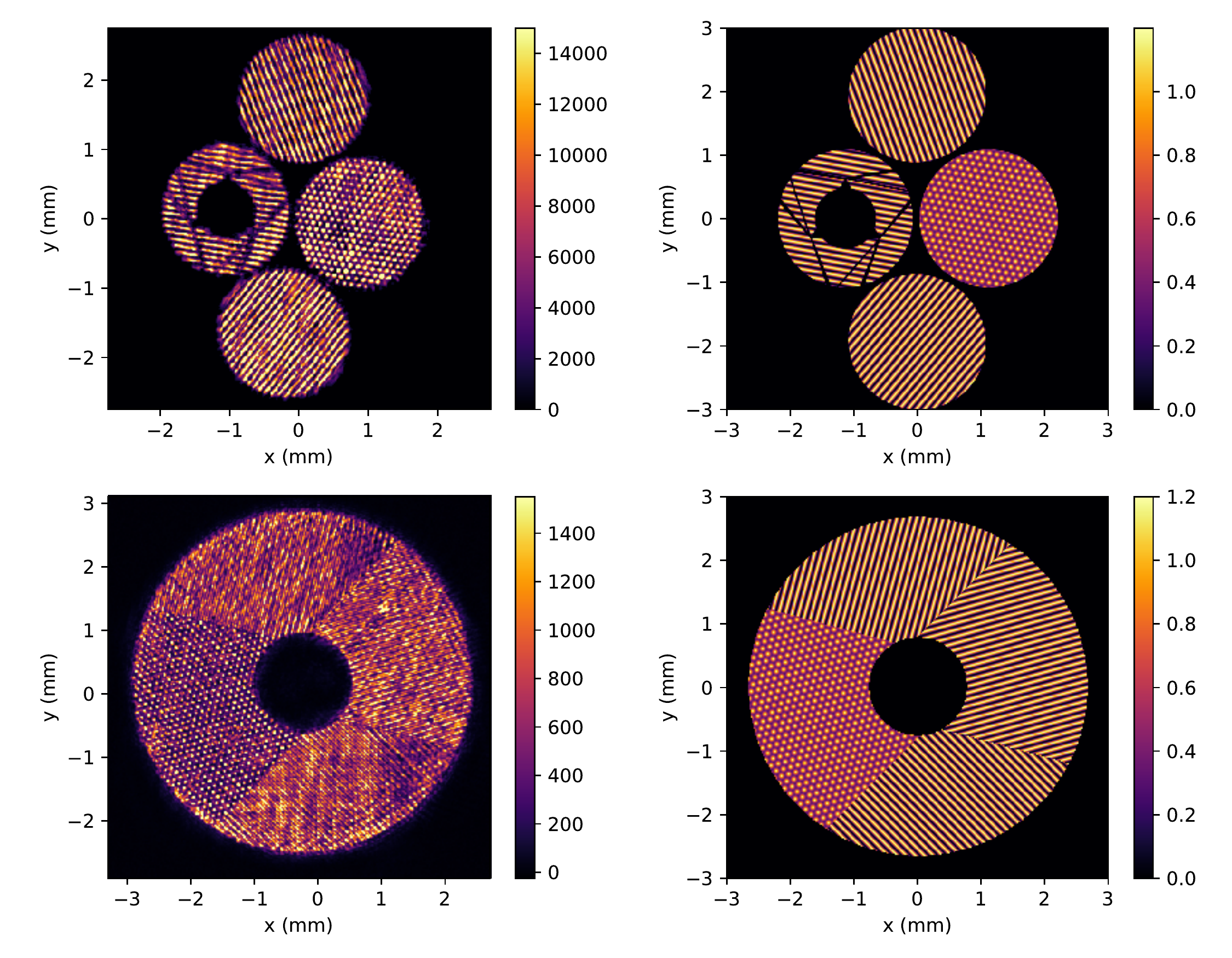}
 \caption{The left column shows the measured pupil images for the P-HCAT mode \edited{with 4 of the 7 GMT segments (one of which can be moved in piston)} (top) and for MagAO-X's annular pupil (bottom). The column on the right shows the theoretical design projected on top of the apertures.}
 \label{fig:phcat_hdfs}
\end{figure*}

The HDFS was tested with the Magellan Adaptive Optics eXtreme (MagAO-X) system \cite{males2018magao, close2018optical}. MagAO-X is a new direct imaging instrument specifically developed for visible high-contrast imaging, a schematic of the system is shown in Figure \ref{fig:magaox}. MagAO-X uses a woofer-tweeter architecture with an ALPAO-97 DM as woofer and a Boston Micromachines 2K tweeter. The system accepts an f/11 beam that first hits the woofer and then the tweeter.  This beam is then relayed to the lower bench where the pyramid wavefront sensor (PWFS) and the science cameras sit. MagAO-X is also used as the AO instrument for the HCAT testbed. HCAT is the GMT High-Contrast Adaptive optics phasing Testbed (HCAT)\cite{hedglen2022hcat}. HCAT simulates the full GMT aperture, where every segment can be actuated in tip/tilt and piston. The beam is then fed into MagAO-X, which allows MagAO-X to operate as if it is an AO instrument for the GMT. Together, HCAT and MagAO-X will be used to test various techniques and algorithms for segment phasing for the GMT\cite{hedglen2022hcat}. Currently, a prototype HCAT (P-HCAT) simulates 4 of the 7 GMT segments, with 1 segment fully controllable in tip, tilt and piston. The prototype testbed was rapidly fabricated to quickly mature closed-loop piston control of the PWFS, and test the injection into MagAO-X. However, it was also perfect to utilize for the first tests of the HDFS.
This section describes the first results of closed-loop piston control with and without turbulence using P-HCAT/MagAO-X and the HDFS. The HDFS was manufactured by Beam Co. and uses the geometric phase property of liquid crystals to create achromatic phase patterns \cite{komanduri2013multi, escuti2016controlling}. \edited{The manufactured HDFS pattern is not optimal in a SNR sense. For the lab experiments we decided to use a single segment as the phase reference instead of the chain method. This design choice was driven by the fact that only 1 mirror segment could be controlled in piston. And the design uses} a continuous phase instead of the binary phase. The more optimal \edited{binary} designs that are discussed in the theoretical section of this work were \edited{found} after the manufacturing of the initial HDFS patterns. \edited{However, the operational principle and general behavior of the manufactured HDFS and the optimized HDFS patters are the same.} The pattern together with pupil images inside the instrument are shown in Figure \ref{fig:phcat_hdfs}.

The HDFS was tested with both the P-HCAT and MagAO-X aperture, because P-HCAT has similar behavior to the GMT aperture and MagAO-X has \edited{aperture segment shapes that are similar to the petal modes of the E-ELT/TMT. Testing the HDFS with both apertures will show that it can
phase the GMT and control the petal modes of the E-ELT/TMT.} We use the MagAO-X tweeter to introduce piston in the system. While the tweeter can mimic petal modes well, there is a limited dynamic range of $\pm 500$ nm of piston in surface error. P-HCAT has a dedicated PI S-325 stage for the piston and tip/tilt control of the center right segment (see Figure \ref{fig:phcat_hdfs}). The PI stage has a dynamic range of $\pm 30$ microns piston OPD, which allow us to also test the large capture range of the HDFS.

\subsection{Empirical calibration of the HDFS}
The HDFS is calibrated by going through a range of piston amplitudes, similar to the template matching method. However, instead of using simulated fringes for the reconstruction, we are using the measured fringes. While this approach allow us to accurately calibrate the HDFS response, it will not drive the closed-loop control to zero piston. The closed-loop control will drive us towards zero piston according to the calibration images, which may or may not have a small piston offset. This bias can be removed by using a simulated set of templates. However, this requires fitting an instrument model, which is not within the scope of the current work. We phased the instrument by first optimizing the HDFS fringe shape and then \edited{removed NCPA by} optimizing the peak flux of the PSF. \edited{After removing the NCPA, we optimized the HDFS fringe shape again because the initial phasing step could have been effected by the NCPA optimization.} This sequence brings the instrument very close to zero differential phase across all segments (white light fringe condition). \edited{The phasing calibration} is only relevant for the GMT because the MagAO-X aperture does not have phasing issues since it is a continuous aperture.

\begin{figure*}[htbp]
 \centering
 \includegraphics[width=\textwidth]{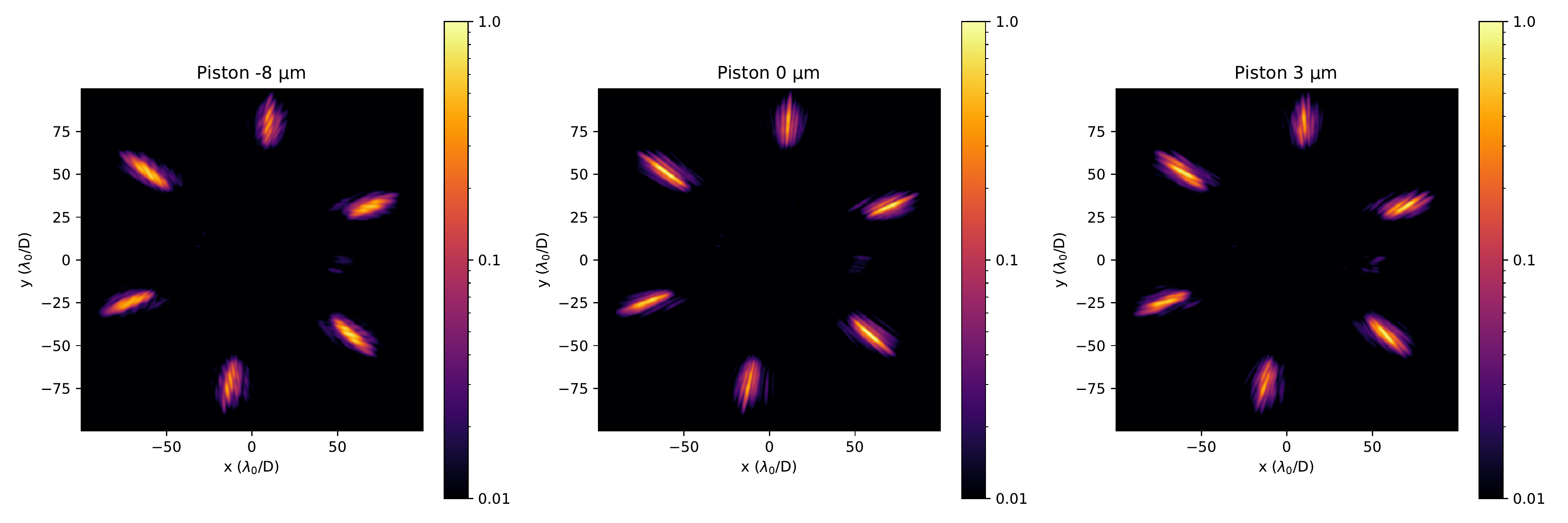}
 \caption{The focal plane images from P-HCAT with different amounts of piston applied to the actuated segment. The central PSF is numerically masked to enhance the dynamic range of the image. The centre image shows the phased system with straight fringes. And, the images on the left and right show the response of the HDFS to piston offsets \edited{(shown on top of each figure in amount of total wavefront error)}. The fringes create the typical barber pole shape. The amount of wrapping in the fringe depends linearly on how many times the phase wraps.}
 \label{fig:calibration_images}
\end{figure*}

Three examples of the template library for P-HCAT are shown in Figure \ref{fig:calibration_images}. The figures are shown for several piston offsets. The fringes wrap around for large amount of piston and show the typical "barber pole" shape. All fringes react to a piston offset because this particular design interferes all segments (or petals) with a single reference segment. Each fringe will react if that one segment is actuated.

\begin{figure*}[htbp]
 \centering
 \includegraphics[width=\textwidth]{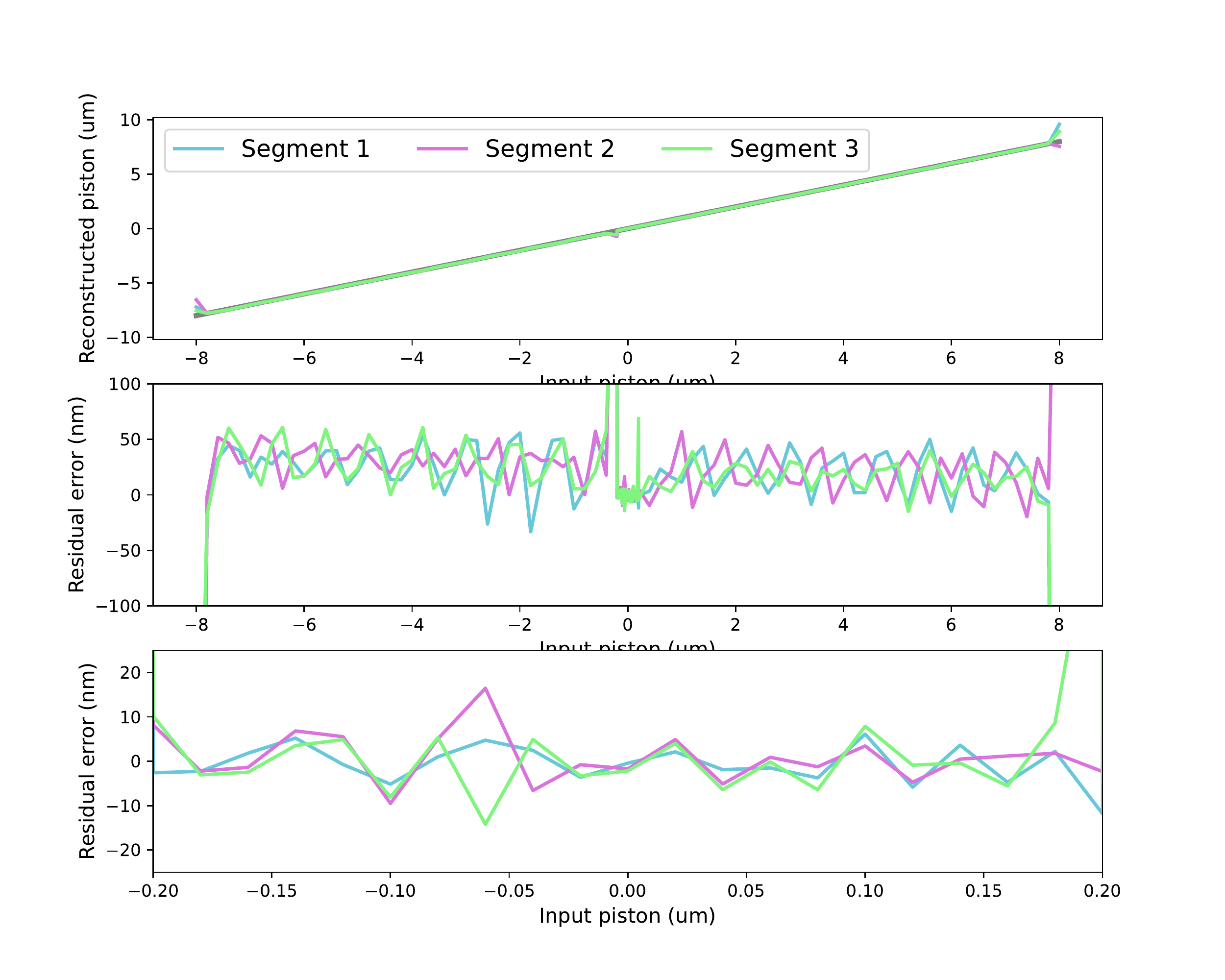}
 \caption{The reconstruction quality of the empirical calibration of the HDFS for different segments of the MagAO-X pupil. The top panel shows the reconstructed piston versus the input piston over a range of $\pm 8 \mathrm{\mu m}$. The middle panel shows the residual piston error. And the lower panel is a zoom in of the residual error in the well calibrated regime. The edge effects at the extremes ($\pm 8 \mathrm{\mu m}$) and the transition region ($\pm 200$ nm) are from the second-order polynomial fitting step.}
 \label{fig:cal}
\end{figure*}

\edited{The quality of the empirical calibration is shown in Figure \ref{fig:cal}. The HDFS has been calibrated in steps of 200 nm for large piston values ($>$ 200 nm) and in steps of 20 nm for small piston values ($<$ 200 nm). The reconstruction shows that the template matching is more accurate with smaller steps between the templates. However, the accuracy is not necessary over the full dynamic range if the system operates in closed-loop. If a large piston error is present then the closed-loop operation will bring it into the well calibrated regime ($<$200 nm). At that point the better calibration takes over and even smaller residuals are possible. The rms of the residual open-loop errors is 5 nm in the small piston regime. The quality of the reconstruction is determined by the dynamical stability of MagAO-X and P-HCAT. Any dynamical error will become part of the template library. Averaging several measurements gets rid of the high-frequency dynamics, but not the slowly evolving ones.}

\subsection{MagAO-X closed-loop results}
\begin{figure*}[htbp]
 \centering
 \includegraphics[width=\textwidth]{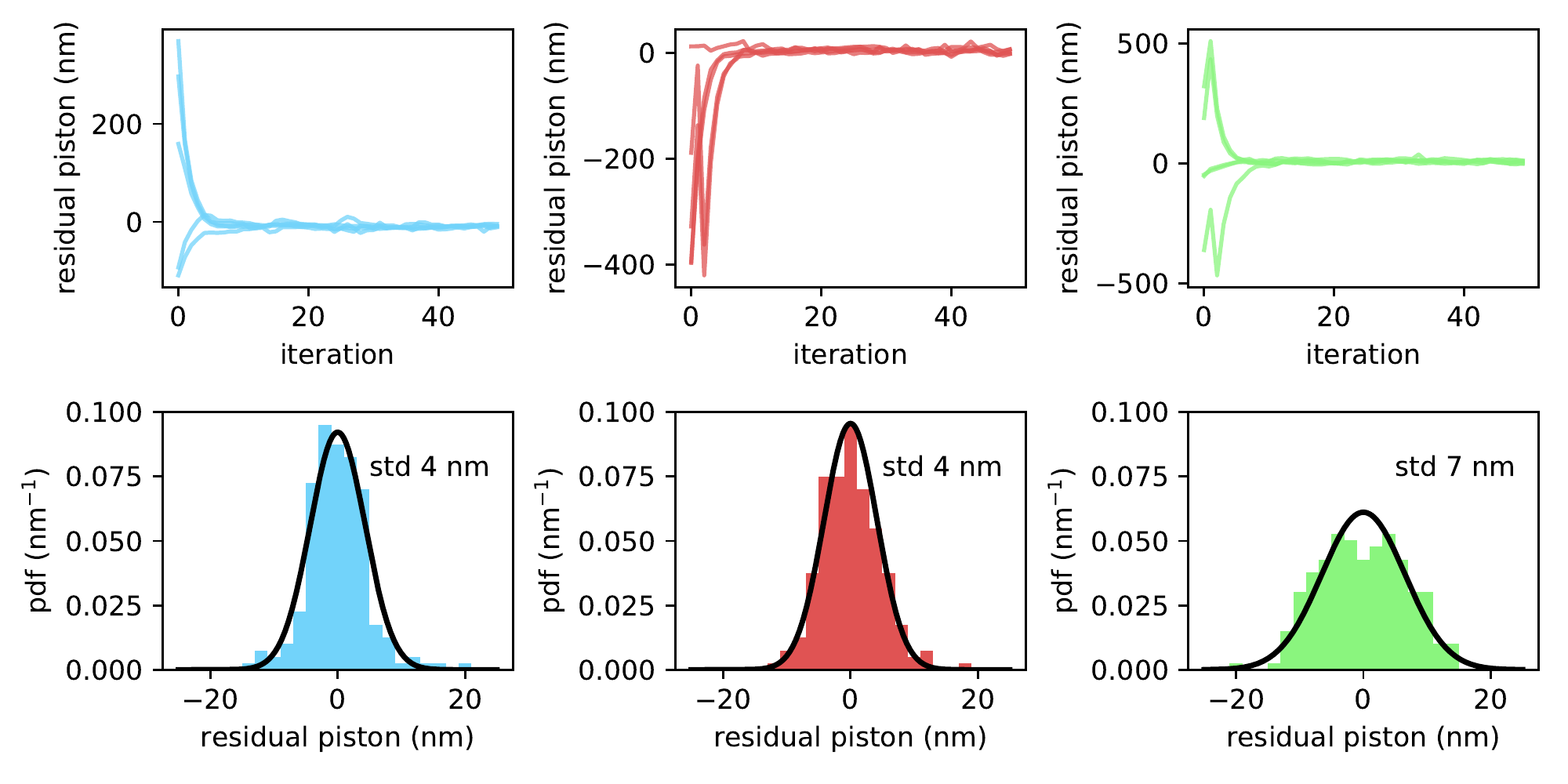}
 \caption{The top row shows the closed-loop residuals for the three different petals of the MagAO-X pupil. Each line shows a different trial run. The bottom row shows the histogram of all residuals after iteration 20. A Gaussian distribution matching the standard deviation of each histogram is plotted for each histogram. The residuals are at the nm level.}
 \label{fig:magaox_closed_loop}
\end{figure*}

All petals can be controlled in piston with the tweeter of MagAO-X. This allowed us to test simultaneous control of multiple segments. The results of the closed-loop tests are shown in Figure \ref{fig:magaox_closed_loop}. We ran several trials, and at the start of each trial a random offset was added to the petal. The system was then run in closed-loop with a gain of 0.3 to remove the piston. The HDFS was able to remove all applied piston to an rms level of several nm. This shows that the HDFS can be used as a very accurate piston sensor.

\begin{figure*}[htbp]
 \centering
 \includegraphics[width=\textwidth]{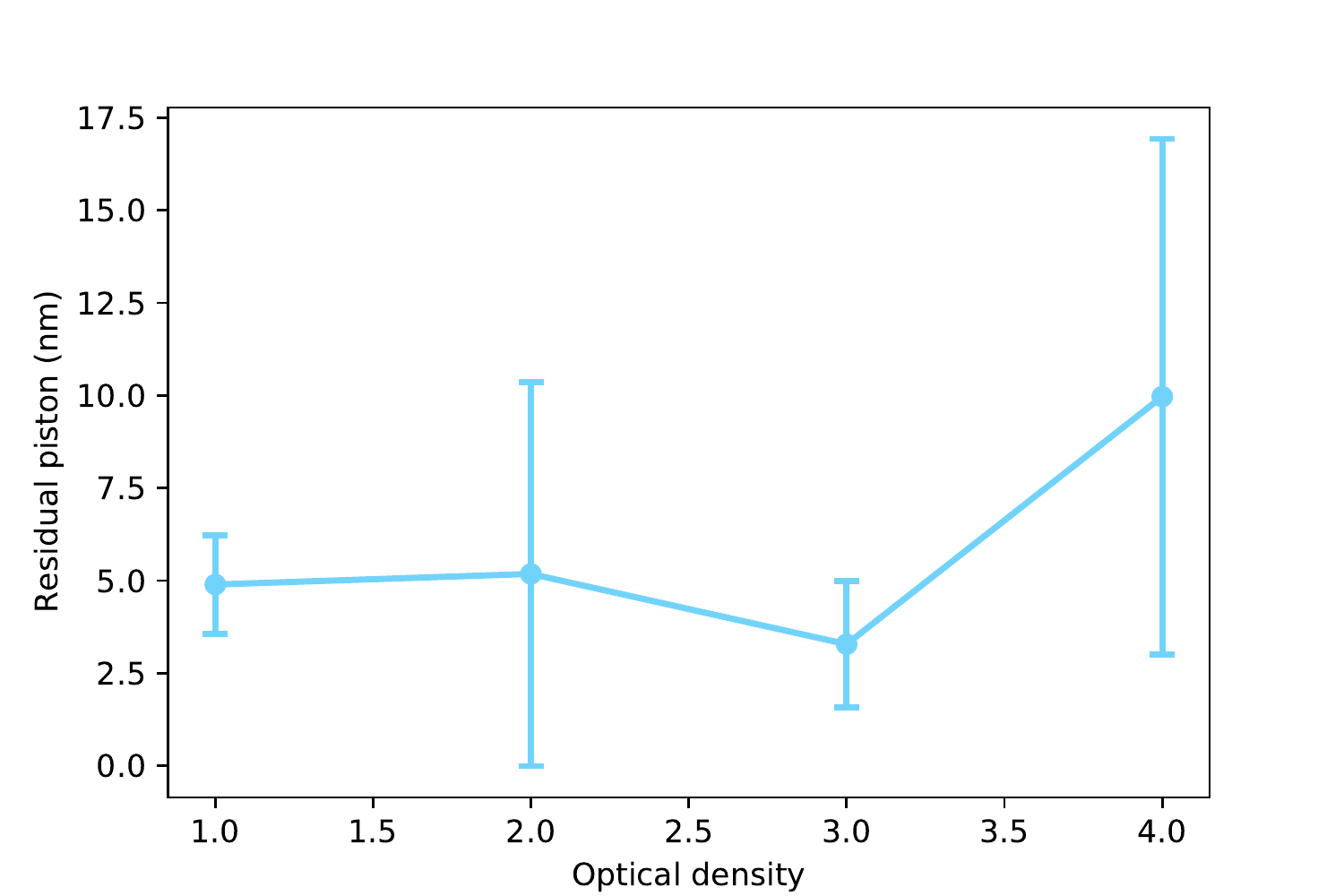}
 \caption{The average rms as a function of filter optical density. The residual piston signal starts to degrade at an optical density of 4, which is $10^{-3}$ times fainter than an optical density of 1. For all optical densities the HDFS reaches a residual piston rms of 10 nm or better.}
 \label{fig:hdfs_fainter}
\end{figure*}

We have also tested the behavior as function of photon flux by switching to different ND filters (see Figure \ref{fig:hdfs_fainter}). The closed-loop performance was almost independent of the photon flux. There was only a slight degradation in performance for the strongest ND filter. The photon flux of the MagAO-X source without any ND filter in it, corresponds to about a zeroth magnitude star. Therefore, an ND 4 filter simulates a 10th magnitude star. The performance is most likely limited by the \edited{calibration of the templates. The rms of the closed-loop residuals is almost identical to the rms of the open-loop reconstruction.} 

\subsection{P-HCAT closed-loop results}
P-HCAT injects its beam into MagAO-X from a separate optical table. While this table is floated on air, it is not actively controlled like the optical table of MagAO-X. Secondly, the optical path length of P-HCAT is quite long and it is not baffled. This creates some low-order turbulence. One of the main challenges for P-HCAT was the calibration of the HDFS sensor. We had to run the PWFS of MagAO-X in closed-loop to control the low-order turbulence to create a stable enough PSF to make the template library. Special care was taken to make sure that the PWFS was not correcting or inducing any piston with the 2k Tweeter DM.

\begin{figure*}[htbp]
 \centering
 \includegraphics[width=\textwidth]{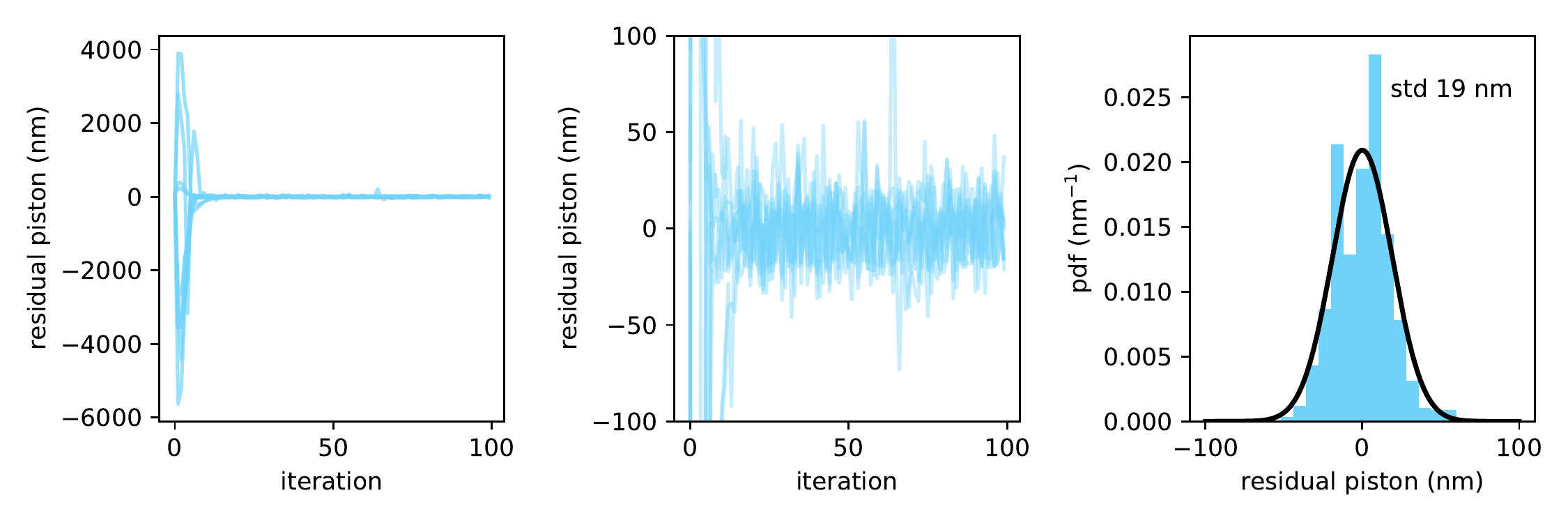}
 \caption{Closed-loop performance of the HDFS with the P-HCAT aperture without any injected disturbance. The closed-loop residuals are shown in the left panel with a zoom-in showed in the middle panel. A large piston offset is injected before the control-loop starts. The distribution of the residuals is shown in the histogram on the right. The HDFS reaches a rms of 19 nm. }
 \label{fig:phcat_cl_noatmos}
\end{figure*}

The stability and precision of the HDFS was tested by perturbing the PI stage and running the system in closed-loop. The results for several trial runs are shown in Figure \ref{fig:phcat_cl_noatmos}. The initial offsets are not several hundred nm, but several microns of piston which would completely confuse any phase wrapping piston sensor (like a PWFS or a phase diversity sensor). The HDFS could close the loop regardless of the amount of piston we applied and, reached a final rms of about 19 nm. The increase from 4-7 nm with MagAO-X to 19 nm for P-HCAT can be attributed to small miscalibrations of the template library because each measurement still has some low-order residual bench turbulence, even after control with the PWFS. 

\begin{figure*}[htbp]
 \centering
 \includegraphics[width=\textwidth]{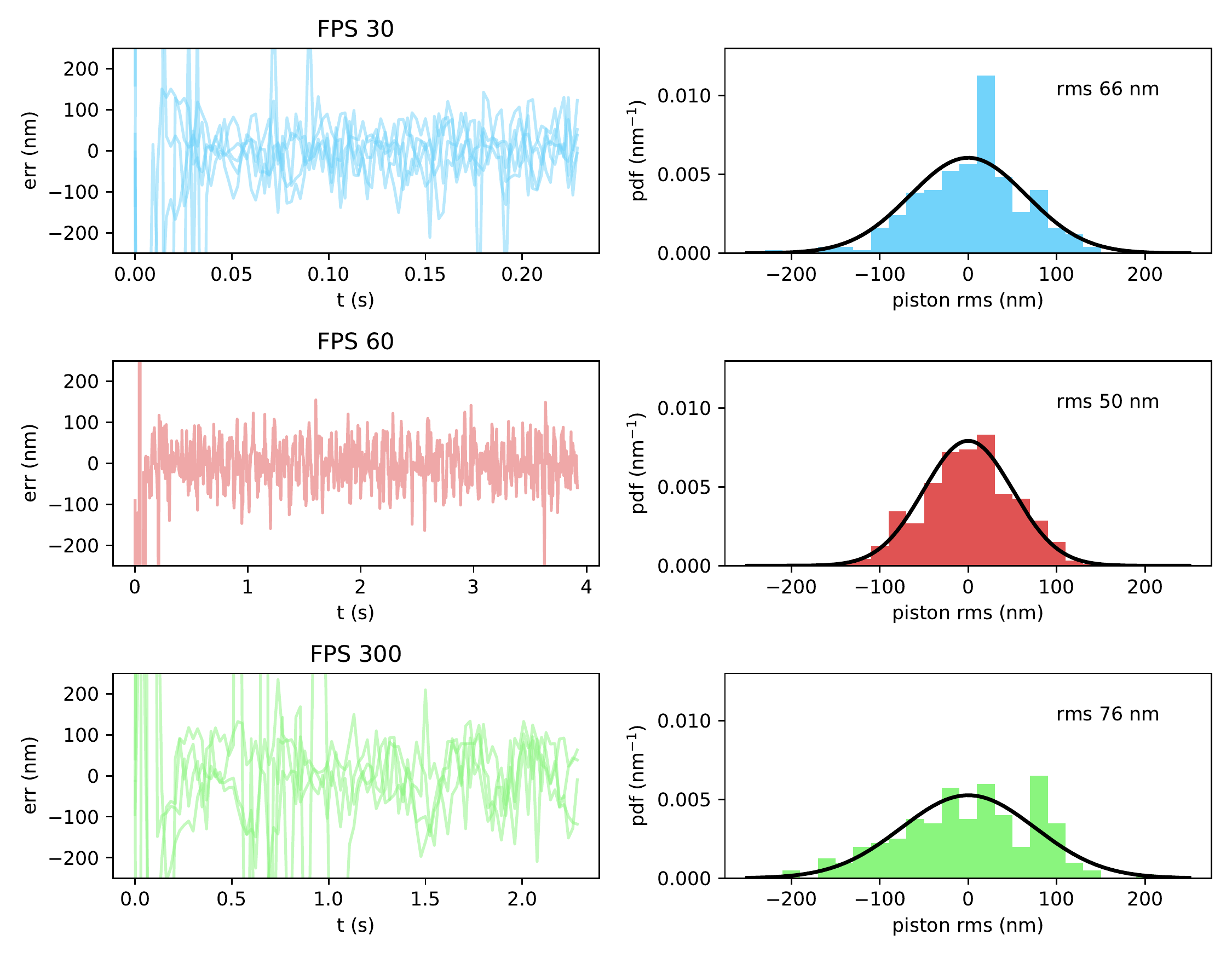}
 \caption{The closed-loop residuals for different AO loop speeds. At low loop speeds (30 FPS), the bench turbulence is not well corrected which increases the HDFS residuals to 66 nm. At high AO loop speeds (300 FPS), the injected turbulence changes too fast and the HDFS can not keep up, which results in a higher rms of 76 nm. The optimum is around 60 FPS with a closed-loop rms of 50 nm. }
 \label{fig:phcat_hdfs_cl}
\end{figure*}

We also investigated the effect that residual turbulence has on the closed-loop performance of the HDFS. Therefore, 0.65 arcsec seeing turbulence was generated with the MagAO-X tweeter DM. This corresponds to the median seeing conditions at Las Campanas Observatory \cite{thomas2010giant}. The PWFS was used purely as a slope sensor to correct the wavefront errors excluding piston, while the HDFS took complete control of the piezo segment. After generating turbulence with the tweeter DM and closing the loop with the PWFS, we again injected random amounts of piston and closed the loop with the HDFS. The results are shown in Figure \ref{fig:phcat_hdfs_cl}. In this way the HDFS was able to successfully close the loop with the piezo segment from p-HCAT for median seeing conditions with about 50\,nm\,RMS.

\edited{The main limitation of the closed-loop control in the p-HCAT mode is read-out speed of the science camera.} Since the MagAO-X science cameras only run as fast as 20\,Hz, we could only run the HDFS at 20\,Hz, which was too slow to keep up with the simulated atmosphere that was injected at 300 Hz. To improve the control, we need a faster camera in the focal plane. In 2022 we will incorporate a faster camera and repeat these tests, but for now the way we overcame this issue was by slowing down the turbulence on the DM. The PWFS was still operated at high speed (300\,Hz) to control the bench seeing from the p-HCAT lab, but the injected atmospheric turbulence and correction were updated at 30\,Hz, 60\,Hz and 300\,Hz to simulate different relative loops speeds of the HDFS. Running the turbulence at such frequency means that the HDFS was effectively running at, 433\,Hz, 217\,Hz and 43\,Hz respectively. Figure \ref{fig:phcat_hdfs_cl} shows a degradation of the performance at the slow and fast atmospheric tests. At slow speeds, the PWFS is only controlling tip and tilt at the full framerate of the PWFS (300 Hz). Therefore, the other low-order modes from the bench turbulence are dominating the HDFS reconstruction errors. At the other end, the HDFS is effectively running at a low frequency compared to the evolution of the atmosphere. There is an increase in the residuals because the \edited{controller is not able to keep up with the changes in piston induced by the atmospheric turbulence and we build up servo-lag error.} The optimum is found at 60 FPS, where the HDFS reaches a rms of 50 nm. The results can be improved by using a higher speed camera for the HDFS.

\section{Conclusion}
In this paper, we presented the HDFS, a new approach for sensing segment piston and/or petal piston for the GSMTs. We showed its performance both in simulation and with lab experiments. The HDFS uses holography to interfere individual segments of the aperture on separate spatial locations in the focal plane. The HDFS inherently includes dispersion due to the diffraction properties of holography. The dispersed fringes have a large dynamic range if it is used over a broad wavelength range. This results in a sensor that is both very sensitive to piston \edited{($<$ 10 nm rms)} and has a large dynamic range ($>\pm$10 um). We showed the design procedure for the HDFS and investigated how to optimally concentrate the light into the dispersed fringes with binary multiplexed gratings. A binary hologram also leads to polarization insensitive phase holograms if implemented with geometric phase optics.

The properties of the HDFS were explored through numerical simulations for the GMT and E-ELT/TMT apertures. From this we found that a 12.5th (J+H) magnitude star or brighter at GMT can reach an rms of 10 nm for piston control, while for the E-ELT a guide star magnitude of 14 reaches 10 nm rms. The reconstruction rms does depend on the stability of the low-order modes. The rms increases linearly with low-order wavefront rms. To achieve 10 nm or better rms reconstruction of piston requires 10 nm or better control of the low-order modes. In closed-loop with AO residuals, the HDFS reaches 5 nm rms independent of guide star magnitude, which implies that the residuals are limited by the dynamics of the controller.

The lab experiments confirmed the sensitivity of the HDFS. We successfully closed-loop the loop and controlled piston to the nm level. The sensor achieved roughly 5 nm rms of differential piston in closed-loop with the MagAO-X tweeter. On the P-HCAT aperture, the HDFS reached an rms of 19 nm. And, importantly, we reached an rms of 50 nm under median seeing conditions (0.65" at Las Campanas Observatory) with the PWFS running concurrently as a slope sensor in closed-loop. These lab experiments demonstrate the architecture we propose for any of the GMT/TMT/E-ELT, with the PWFS as a slope sensor and the HDFS as a second channel piston\edited{/petal} sensor.

An important point for the HDFS is that it requires good AO correction for each segment. The interference fringes require a stable Airy core in the PSF. Such quality may not always be available in the visible wavelength range. Therefore, it is recommended to use the HDFS in the NIR where the Strehl is intrinsically higher than in the visible. 

Both the numerical and the experimental results show that the HDFS is an excellent piston/petal sensor. It is currently also \edited{chosen as} the second channel piston sensor for the GMT. P-HCAT will be upgraded to HCAT next year \edited{(2023)}, after which it will include a full GMT simulator where all 7 segments can be actuated and controlled\cite{hedglen2022hcat}. The HDFS optic that can work with all 7 segments of HCAT has already been manufactured. Our next step will be to explore piston control of all 7 segments of the GMT testbed simulator together with the PWFS running in closed loop. The imaging cameras of MagAO-X will also be replaced during the HCAT upgrade. The new cameras will be able to run at a much higher framerate. This will allow us to test high-speed piston control. At the same time, we are integrating the HDFS into the simulation environment of the GMT AO system to explore its performance in the relevant parameter space. This will lead to a final design of the HDFS for the GMT.

\subsection*{Disclosures}
The authors declare that they have no relevant financial interests in the manuscript and no other potential conflicts of interest to disclose.

\subsection* {Acknowledgments}
Support for this work was provided by NASA through the NASA Hubble Fellowship grant \#HST-HF2-51436.001-A awarded by the Space Telescope Science Institute, which is operated by the Association of Universities for Research in Astronomy, Incorporated, under NASA contract NAS5-26555. The HCAT testbed program is supported by a NSF/AURA/GMTO risk-reduction program contract to the University of Arizona (GMT-CON-04535, Task Order No. D3 High Contrast Testbed (HCAT), PI Laird Close). Alex Hedglen received a University of Arizona Graduate and Professional Student Council Research and Project Grant in February 2020, which helped provide funds for the Holey Mirror for p-HCAT. Alex Hedglen and Laird Close were also partially supported by NASA eXoplanet Research Program (XRP) grants 80NSSC18K0441 and 80NSSC21K0397 and the Arizona TRIF/University of Arizona "student link" program. We are very grateful for support from the NSF MRI Award \#1625441 (for MagAO-X) and funds for the GMagAO-X CoDR from the University of Arizona Space Institute (PI Jared Males) as well. \edited{The NSF award to AURA is "Cooperative Support Award  \#2013059", and the AURA subaward to GMTO is NE0651C.}


\bibliography{references}   
\bibliographystyle{spiejour}   


\vspace{2ex}\noindent\textbf{Sebastiaan Y. Haffert} is a NASA Hubble Postdoctoral Fellow at the University of Arizona's Steward Observatory. His research focuses on high-spatial and high-spectral resolution instrumentation for exoplanet characterization.

\vspace{2ex}\noindent\textbf{Jared R. Males} is an Assistant Astronomer in the the University of Arizona's Steward Observatory and is the Principal Investigator of MagAO-X.  His research is focused on using high-contrast imaging to study extrasolar planets.

\vspace{2ex}\noindent\textbf{Alexander D. Hedglen} is an Optical Sciences PhD student at the University of Arizona. He received his BA in Astronomy and BS in Physics from the University of Hawai`i at Hilo in 2016. His research focuses on adaptive optics for the direct imaging of exoplanets.

\vspace{1ex}
\noindent Biographies and photographs of the other authors are not available.

\listoffigures
\listoftables

\end{spacing}
\end{document}